\documentclass[%
reprint,
english,
superscriptaddress,
showpacs,preprintnumbers,
amsmath,amssymb,
aps,
prl,
]{revtex4-1}

\usepackage{bm} % bold math
\usepackage[unicode=true,
bookmarks=true,bookmarksnumbered=false,bookmarksopen=false,
breaklinks=false,pdfborder={0 0 1},backref=false,colorlinks=false]
{hyperref}
\usepackage[T1]{fontenc}
\usepackage[utf8]{inputenc}
\usepackage{babel}
\usepackage{float}
\usepackage{units} %for \newboolean
\usepackage{multirow}
\usepackage{sansmath}
\usepackage{amssymb}
\usepackage{graphicx}  
\usepackage{breakurl}  %linebreak in URLs
\usepackage{libertine} 
\usepackage{siunitx}

\usepackage[usenames, dvipsnames]{color}
\pdfoutput=1

\hypersetup{pdftitle={Correlating decoherence in transmon qubits: Low frequency noise by single fluctuators}, pdfauthor={Steffen Schlör}}

\usepackage{color}

\binoppenalty=\maxdimen
\relpenalty=\maxdimen

\newcommand{\angstrom}{\mbox{\normalfont\AA}}

\usepackage{ulem} % required for \sout - strikeout text
\begin{document}
	
	\title{Correlating decoherence in transmon qubits:\\
		Low frequency noise by single fluctuators}
	
	\author{Steffen Schl\"or}   
	\email{Steffen.Schloer@kit.edu}
	\affiliation{Institute of Physics, Karlsruhe Institute of Technology, 76131 Karlsruhe, Germany}
	\author{J\"urgen Lisenfeld}
	\affiliation{Institute of Physics, Karlsruhe Institute of Technology, 76131 Karlsruhe, Germany}
	\author{Clemens M\"uller}    
	\affiliation{IBM Research Z\"urich, 8803 R\"uschlikon, Switzerland}
	\affiliation{Institute for Theoretical Physics, ETH Zürich, 8092 Zürich, Switzerland}
	\author{Alexander Bilmes}
	\affiliation{Institute of Physics, Karlsruhe Institute of Technology, 76131 Karlsruhe, Germany}
	\author{Andre Schneider}    
	\affiliation{Institute of Physics, Karlsruhe Institute of Technology, 76131 Karlsruhe, Germany}
	\author{David P. Pappas}    
	\affiliation{National Institute of Standards and Technology, Boulder, CO 80305, USA}
	\author{Alexey V. Ustinov}
	\affiliation{Institute of Physics, Karlsruhe Institute of Technology, 76131 Karlsruhe, Germany}
	\affiliation{Russian Quantum Center, National University of Science and Technology MISIS, 119049 Moscow, Russia}
	\author{Martin Weides}
	\email{Martin.Weides@glasgow.ac.uk}
	\affiliation{Institute of Physics, Karlsruhe Institute of Technology, 76131 Karlsruhe, Germany}
	\affiliation{James Watt School of Engineering, University of Glasgow, Glasgow G12 8LT, UK}

	\date{\today}  % always today, but any date may be explicitly specified

	\begin{abstract}
		We report on long-term measurements of a highly coherent, non-tunable superconducting transmon qubit, revealing low-frequency burst noise in coherence times and qubit transition frequency. We achieve this through a simultaneous measurement of the qubit's relaxation and dephasing rate as well as its resonance frequency. The analysis of correlations between these parameters yields information about the microscopic origin of the intrinsic decoherence mechanisms in Josephson qubits. Our results are consistent with a small number of microscopic two-level systems located at the edges of the superconducting film, which is further confirmed by a spectral noise analysis.
	\end{abstract}
	
	\keywords{superconducting qubits, coherence, noise, two-level system, correlation, PSD}
	
	\maketitle
	%\section{\label{sec:level1}Introduction:}
	Today\textquoteright s prototype solid-state quantum computers built from superconducting qubits such as the transmon~\cite{Koch2007} 
	are already capable of finding the electronic ground state of small molecules~\cite{Kandala2017}. 
	Their complexity keeps growing, while error rates of logical gate operations are already close to the threshold for some fault-tolerant quantum computing schemes~\cite{Barends2014,Fowler2012}. 
	However, the error probability due to random parameter fluctuations scales exponentially with the number of qubits, rendering the calibration of many-qubit systems difficult. 
	The demand on stability and coherence of scaled-up quantum systems widens the focus of current research towards new decoherence mechanisms 
	and fluctuations occurring on time scales of hours or even days.
	
\begin{figure*}[htb]
		\includegraphics[width=1\textwidth]{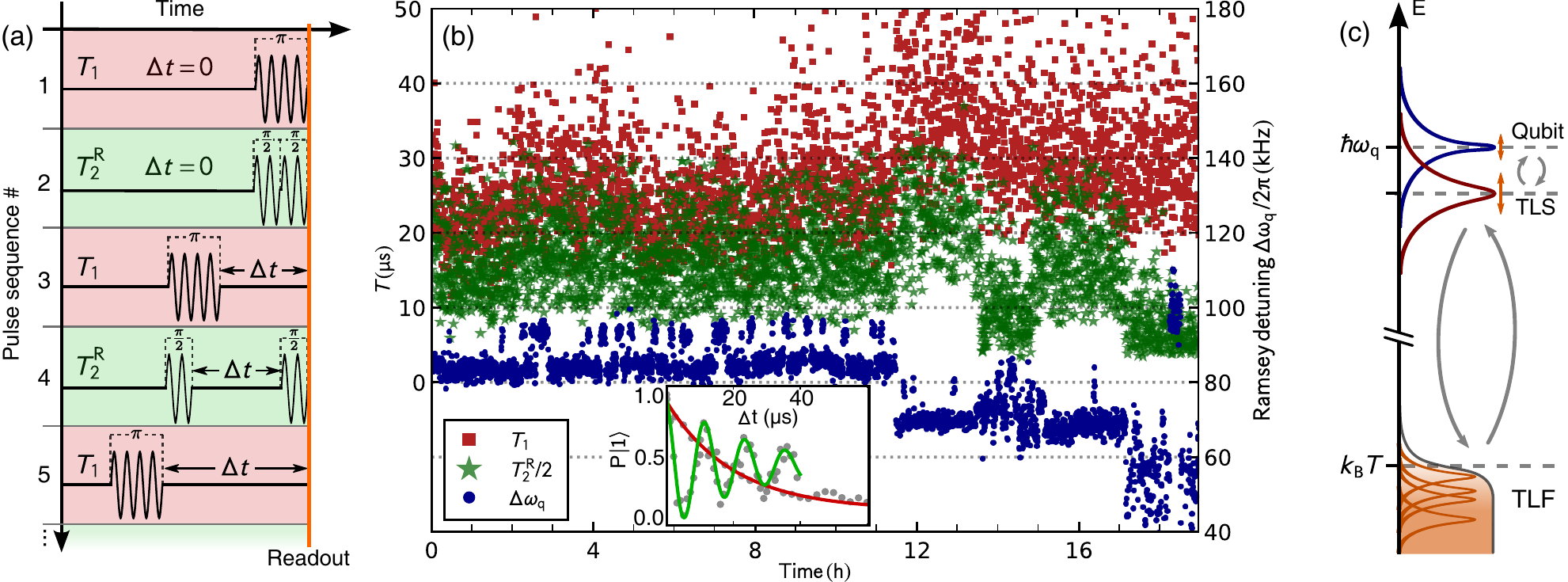}\protect\caption{
			\label{fig:long_measurement1-1} (a) Measurement pattern: Single pulse sequences of different measurements (e.g.\ of $T_{1}$ and $T^\mathrm{R}_{2}$) are interleaved, 
			resulting in a simultaneous acquisition. The time $\Delta t$ is the free evolution. The ratio between the number of pulses can differ, and spin echo pulses may be added. 
			The inset in (b) shows an exemplary single trace with fits for $T_{1}$ (red), $T_{2}^\mathrm{R}$ and the Ramsey detuning $\Delta\omega_{\mathrm{q}}$ (green). 
			(b) Data taken over a course of 19 hours displays fluctuations in $T_{1}$ and $T_{2}^{\mathrm{R}}$ (red squares and green stars, left axis), 
			and telegraph-like switching of the qubit frequency $\Delta \omega_{\mathrm{q}}$ (blue dots, right axis). 
			The time resolution corresponds to  $10\,\mathrm{s}$ of averaging using the pattern shown in (a).
			For clarity, dephasing times are divided by two. With this measurement we reveal a connection between noise at mHz frequencies and qubit dephasing.  
			(c) Illustration how the frequency of a single TLS near resonance with the qubit fluctuates due to its coupling to thermally activated TLS (so-called TLF) at energies at 
			or below $k_\mathrm{B}\,T$ (orange shaded area). 
			Depending on the detuning between qubit and TLS, this can cause positive or negative correlations between qubit coherence times and its resonance frequency.
		}
	\end{figure*}

	To examine the stability of a transmon-type qubit, we perform long-term measurements of energy relaxation $T_1$, Ramsey $T_2^\mathrm{R}$, and spin echo $T_{2}^{\mathrm{E}}$ coherence times, as well as the transition frequency $\omega_\mathrm{q}$. 
	When these parameters are measured consecutively, inconsistencies are possible due to fluctuations.
	Here, we develop and employ a time-multiplexed pulse sequence pattern (see Fig.~\ref{fig:long_measurement1-1} (a)) which allows us to acquire all qubit parameters simultaneously.
	Moreover, the interleaved pattern enables us to characterize correlations of qubit parameter fluctuations and coherence, 
	which reveal a connection between noise at mHz frequencies and qubit dephasing.
	
	Our long-term measurements reveal significant fluctuations in all qubit parameters, similar to earlier reports \cite{Klimov2018,Paik2011,Dial2016}. Figure~\ref{fig:long_measurement1-1}(b) shows exemplary results of a continuous measurement over 19\,hours. The qubit transition frequency displays telegraphic noise with multiple stationary points, which prompts our interpretation of the data in terms of an ensemble of environmental two-level systems (TLS) interacting with the qubit. 
	TLS may emerge from the bistable tunneling of atomic-scale defects~\cite{Anderson1972,Phillips:1987,Mueller2017} 
	which may reside within the amorphous $\mathrm{AlO_x}$ of the qubit's tunnel barrier or electrode surface oxides, but can also be formed by adsorbates or processing residuals on the chip surface~\cite{Holder:PRL:2013,Graaf2018}. 
	Such defects may couple to the qubit by their electric dipole moments, leading to absorption of energy and fluctuations in qubit parameters. The TLS' parameters are broadly distributed and those TLS having transition frequencies near or at the qubit's resonance can cause dispersive frequency shifts~\cite{Faoro2015}, avoided level-crossings~\cite{Martinis2005, Weides2011, Weides2011a} or resonances in qubit loss~\cite{Klimov2018}. 
	
	%First evidence that TLS are a dominating decoherence mechanism for superconducting qubits was reported by Martinis \textit{et al}.~\cite{Martinis2005}, leading to improved qubit designs where the number of TLS is reduced by smaller tunnel junctions~\cite{Steffen:PRL:2006}. Moreover, capacitive circuit components were optimized to reduce electric fields~\cite{Wang:APL:2015} and hence the coupling to TLS. \com{skipped due to length limitation}
	
	We attribute the observed fluctuations in qubit parameters to a sparse ensemble of environmental TLS close to the superconducting film edge and its interaction with thermal fluctuators. This model is supported by the power spectral density (PSD) of the observed frequency fluctuations.  
		%Variations in qubit energy $\omega_\mathrm{q}$ at fluctuation frequencies of $0.03-40\,\mathrm{mHz}$ show significant deviations from $1/f$ noise, and are well modeled by a Cauchy (Lorentzian) distribution, further supporting the TLS model.
	Complemented by a cross-correlation analysis, our data provides evidence for a small number of TLS which dominate dephasing if near-resonant, 
	while the $1/f$ noise background we also observe, may emerge from a bath of more weakly coupled TLS~\cite{Dutta1981}. We conclude that even single TLS on the edges of the superconducting films can dominate decoherence and cause random parameter fluctuations in superconducting qubits.
	We find that other sources of fluctuation, like temperature variations, critical current fluctuations, quasiparticle tunneling, or flux vortices play secondary roles in the presented experiment.

	%End of 'introduction' 
	
	Our interpretation of the data according to the interacting defect model~\cite{Faoro2015,Mueller2015,Mueller2017}, is further motivated by recent experiments, where the thermal switching of individual TLS in $\mathrm{AlO_x}$ Josephson junctions was measured directly~\cite{Meisner2018}. Further, spectral diffusion of TLS was recently observed by monitoring the $T_1$ time of a tunable transmon qubit~\cite{Klimov2018}. Our results confirm the findings that single TLS strongly affect qubit coherence,  independent of flux noise. 
	Here, we complement earlier experiments by simultaneous measurements of dephasing and qubit frequency, as well as their correlations, further supported by spectral analysis at $\mathrm{mHz}$ frequencies.

	In the interacting TLS model, defects may mutually interact electrically or via their response to mechanical strain~\cite{Grabovskij2012,Lisenfeld2015}. 
	If the transition energy of a particular TLS is below or close to the thermal level $k_\mathrm{B}\,T$, 
	it undergoes random, thermally activated state-switching. We call these two-level fluctuators (TLF) to distinguish them from the more coherent TLS~\cite{Mueller2017} with higher transition energies. 
	Longitudinal coupling between TLS and TLF causes telegraphic fluctuation or spectral diffusion~\cite{Black:PRB:1977} of the TLS' resonance frequencies. 
	The resulting time-dependent frequency fluctuation of near-resonant TLS give rise to phase noise of superconducting resonators~\cite{Faoro:PRL:2012} 
	and may also cause the parameter fluctuations of qubits~\cite{Mueller2015}, investigated here. Figure~\ref{fig:long_measurement1-1}(c) illustrates the physical picture.
	
	%($\propto\sigma_{\mathrm{z}}$) 
	
	%\subsection{Setup and Measurement:}
	
	We use a non-tunable transmon qubit with an $\mathrm{Al\text{-}AlO_{x}\text{-}Al}$ junction, shunted by coplanar capacitor films of $40\,\mathrm{nm}$ $\mathrm{TiN}$, capacitively connected to a microstrip readout resonator. The Hamiltonian describing our qubit is well approximated by $H_\mathrm{q} / \hbar = \omega_\mathrm{q}\,a^{\dagger}a - \alpha(a^{\dagger})^2(a)^2,$ where $\omega_\mathrm{q}$ is the splitting between the ground and excited state, $\alpha$ is the anharmonicity, and $a^{\dagger}$ and $a$ are the raising and lowering operators. The qubit transition frequency is $\omega_\mathrm{q}/2\pi=4.75\,\mathrm{GHz}$, and the ratio of Josephson energy to charging energy $E_{J}/E_{C}$ is $78$, leaving it well-protected from charge fluctuations \cite{Koch2007}.% We observed $T_{1}$ times between $10$ and $\SI{80}{\micro\second}$, and $T_{2}^{\mathrm{R}}$ between $5$ and $\SI{80}{\micro\second}$.
	
	%In the following we are interested in the qubit relaxation time $T_1$, Ramsey dephasing time $T_2^{\mathrm{R}}$ and its transition frequency $\omega_\mathrm{q}$.
	Repeated measurements with an interleaved sequence analogous to Fig.~\ref{fig:long_measurement1-1}(a) 
	reveal time-dependent dynamics of the qubit parameters, an example of which is shown in Fig.~\ref{fig:long_measurement1-1}(b). The Ramsey detuning $\Delta\omega_{\mathrm{q}}$ (blue dots) is a direct measure for the shift in qubit frequency, which fluctuates between multiple discrete values and also shows abrupt qualitative changes in fluctuation dynamics. 
	The relaxation time $T_{1}$ (red squares) and Ramsey dephasing time $T_{2}^{\mathrm{R}}$ (green stars) show fluctuations and a clear correlation with $\Delta\omega_{\mathrm{q}}$, which we will evaluate in the following. 
	A single slice of this measurement (see inset in Fig.~\ref{fig:long_measurement1-1}(b)) required averaging for about 10\,s. 
	$T_1$, $T_{2}^{\mathrm{R}}$ and $\Delta \omega_{\mathrm{q}}$ were extracted from fits to single traces. 
	See appendix~B for further details on measurement procedure.

	\begin{figure*}[htb]
		\includegraphics[width=1.0\textwidth]{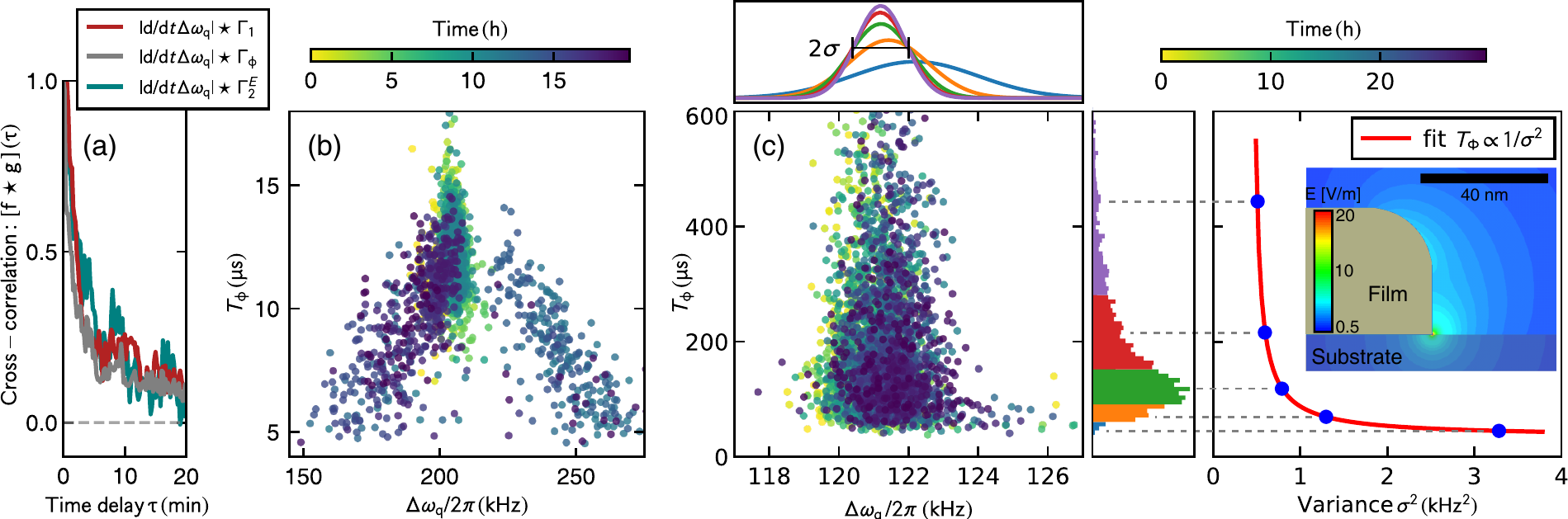}
		\protect\caption{
				\label{fig:scatterplots} (a) Cross-correlation of the absolute fluctuation strength and relaxation or dephasing rates of the dataset shown in (b). All curves show significant correlation at zero time delay $\tau$, relating fluctuations in qubit frequency on the order of seconds to relaxation and dephasing. (b,c) Scatterplots of $T_{\Phi}$ versus Ramsey detuning for two measurements from different cooldowns (identical setup) with drastically different pure dephasing times. The point color indicates the measurement time. In (b) positive, negative, and no correlation occur within the measurement period. In (c) the qubit frequency is relatively stable, bins of pure dephasing times are colored in the vertical histogram, corresponding fits to normal distributions (top panel) are colored accordingly. Lower dephasing times correspond to larger variances in qubit frequency. The standard deviation of the violet curve is indicated exemplarily. On the right, the extracted variances $\sigma^2$ are plotted against the corresponding mean values of pure dephasing. A fit to the expected function $T_\mathrm{\Phi}\propto 1/\sigma^2$ is in agreement with the data. The simulated field distribution at the superconducting film edge of the qubit capacitance is shown in the inset. 
			%which we attribute to frequency drift of a single TLS close to the qubit frequency causing dephasing.
		}
	\end{figure*}

	We describe TLS by the generic two-level Hamiltonian $H_{{\mathrm{TLS}},k}=-\frac{\hbar}{2}(\epsilon_k\sigma_{\mathrm{z}}+\delta_k\sigma_\mathrm{x}),$
	where $k$ is the TLS index, $\epsilon_k$ is the asymmetry energy, $\delta_k$ is the tunneling energy, and $\sigma_i$ are the Pauli matrices. Assuming the standard form of qubit-TLS coupling~\cite{Shnirman2005,Cole2010}  $H_{\mathrm{int},k}=\hbar g_k\,\sigma_\mathrm{z}(a+a^{\dagger}),$ transformation into the dispersive frame yields 
	\begin{equation}
	\begin{split}
	H_\mathrm{q}+H_{\mathrm{TLS},k}+H_{\mathrm{int},k}\approx\hbar(\omega_\mathrm{q}+\chi_{k} \sigma_\mathrm{z})a^\dagger a+ \\ \frac{\hbar}{2}(\omega_{\mathrm{TLS},k}+\chi_{k})	\sigma_\mathrm{z}-\hbar\alpha(a^{\dagger})^2(a)^2,
	\end{split}
	\end{equation}
	where $\chi_{k}=g_{k}^2/\Delta$ is the dispersive shift and the detuning between TLS and qubit is given by $\Delta=\omega_{\mathrm{TLS},k}-\omega_\mathrm{q}$.

	We can estimate the coupling strength $g_k$ between qubit and TLS from the observed fluctuation amplitude $\Delta \omega_{\mathrm{q}}$, 
	assuming resonant TLS with a dipole moment on the order of $1\,\mathrm{e\angstrom}$~\cite{Sarabi2016,Brehm2017,Lisenfeld2016} (see appendix~E for details). 
	The maximum coupling rate, achieved for TLS located in the junction is approximately $\SI{48}{\mega\hertz}$. 
	Such strong coupling would allow for much larger changes in qubit frequency than the observed $5\text{-}140\,\mathrm{kHz}$. 
	By simulating the electric field distribution we find the coupling strength to TLS at sites closer than $20\,\mathrm{nm}$ to capacitor-edges is $g_k\gtrsim100\,\mathrm{kHz}$, in agreement with our observations. Thus we conclude that the dominant TLS in our experiment reside close to film edges. 
	
	%\subsection{Experimental result:}

	To fathom the microscopic origin of the fluctuations, we analyze correlations between all extracted parameters. Ramsey dephasing consists of relaxation and 'pure' dephasing $T_{\mathrm{\Phi}}$, connected by $1/{T_{2}^{\mathrm{R}}}=1/{2T_{1}}+1/{T_{\mathrm{\Phi}}}$. 
	In the following, we focus on $T_{1}$ and $T_{\mathrm{\Phi}}$ or the corresponding rates $\Gamma_{1}=1/T_{1}$ and $\Gamma_{\mathrm{\Phi}}=1/T_{\mathrm{\Phi}}$. Scatterplots of two long-term measurements from successive cooldowns with identical setup are shown in Fig.~\ref{fig:scatterplots}(b) and (c), where $T_{\mathrm{\Phi}}$ is plotted vs. $\Delta \omega_\mathrm{q}$. Fig.~\ref{fig:scatterplots}(b) exhibited generally larger fluctuations and lower dephasing times, different types of correlation could be observed in the course of a single measurement. 
	A time interval of about $10\,\mathrm{h}$ without obvious correlation between $T_{\Phi}$ and $\Delta \omega_\mathrm{q}$ is followed by alternating positive and negative correlation during times of strong frequency fluctuation. Cross-correlation analysis of this data (Fig.~\ref{fig:scatterplots}(a)) relates the absolute fluctuation strength of $\Delta \omega_\mathrm{q}$ to higher dephasing and relaxation rates, linking slow fluctuations on the order of seconds to dephasing or relaxation up to the order of microseconds. We interpret these observations as coupling to a single spectrally diffusing TLS crossing the qubit frequency several times. To our knowledge, no other interpretation is in agreement with our observations, as will be discussed later. The polarity and strength of the correlations depend on the sign of the detuning between TLS and qubit and their mutual coupling strength.

	To perform a quantitative analysis of the connection between the fluctuations in qubit frequency and the pure dephasing time, we examine the variance in qubit frequency associated with multiple ranges of dephasing times (Fig.~\ref{fig:scatterplots}(c)) while the qubit frequency is relatively stable. 
	We bin the frequency shift data according to their associated pure dephasing times, and fit the data in each bin to a Gaussian distribution. Assuming the qubit frequency shifts to be due to random sampling of a linear function (as is the case for small frequency shifts of a dispersively coupled TLS), 
	the standard deviation $\sigma$ of the distributions will be proportional to the slope of this linear function. Conversely, the pure dephasing rate $\Gamma_{\mathrm{\Phi}}$ in such a situation is proportional to the square of the slope of the frequency change with the random parameter~\cite{Slichter2016}. If the origin of the measured large frequency fluctuations is the same as the one for the pure dephasing, we expect the two slopes to be the same, 
	such that for each bin in pure dephasing time we have $\Gamma_{\mathrm{\Phi}} \propto \sigma^{2}$, which is in good agreement with our data.

	In repeated measurements and different cooldowns, we find qubit coherence times to be anti-correlated with the maximum amplitude of frequency fluctuations. 
	In our model, this corresponds to different dispersive shifts $\chi_{k}$ due to the respective dominant TLS. During cooldowns with persistently long relaxation and dephasing times as in Fig.~\ref{fig:scatterplots}(c), this shift is low and qubit frequency fluctuations are small. If increased interaction with a TLS leads to shorter relaxation and dephasing times, even for intermediate times without resolvable frequency fluctuations of the qubit, dephasing tends to stay low. This is expected because of the higher frequency noise we can not resolve by our sub-Hz repetition rate.  
	Possible explanations for abrupt changes in decoherence dynamics are slow thermalization processes in the amorphous parts, logarithmically slow TLS relaxation~\cite{Asban2017}, or background radiation.

	Throughout our measurements, reduced coherence manifests itself most strongly in the dephasing times $T_{\Phi}$ and $T_{2}^\mathrm{R}$ rather than in $T_1$ and spin echo $T_{2}^{\mathrm{E}}$. The observed effective reduction of dephasing by spin echo pulses suggests most of the relevant noise spectrum to lie below the spin-echo cutoff frequency of $25\,\mathrm{kHz}$, in our case. This observation is in agreement with the typical maximum fluctuation rate of thermal TLS due to phonons of $\gamma_{1}^{\mathrm{max}}(T=20\,\mathrm{mK})\approx1.9\,\mathrm{kHz}$~\cite{Lisenfeld2016} in our case.

	%PSD:
	
	To further elucidate the origin of the observed qubit frequency fluctuations, we performed a long term measurement in which we optimized the measurement pulse sequence to gain maximum frequency resolution.
	If the fluctuations are due to individual TLS, we expect the power spectral density to follow the functional form~\cite{Shnirman2005} 
	\begin{equation}
	C(\omega)\propto(1-\langle\sigma_{\mathrm{z}}\rangle ^2)\frac{2 \gamma_{1,k}}{\gamma_{1,k}^2+\omega^2},
	\end{equation} 
	a Lorentz distribution centered at zero frequency. Here, $\gamma_{1,k}$ is the TLS relaxation rate, 
	$\langle\sigma_{\mathrm{z}}\rangle=\tanh(E_k/2k_{\mathrm{B}}T)$ is the thermal equilibrium population of TLS '$k$' and $E_k=\sqrt{\epsilon_k^2+\delta_k^2}$ is its transition energy.
	Under the assumption of a uniform distribution of TLS barrier heights, 
	the superposition of many such Lorentzian spectra are responsible for the typically observed low-frequency noise of the form $\sim1/f^{\alpha}$, 
	usually observed in all solid-state qubits~\cite{Dutta1981}.

\begin{figure}
		\includegraphics[width=\linewidth]{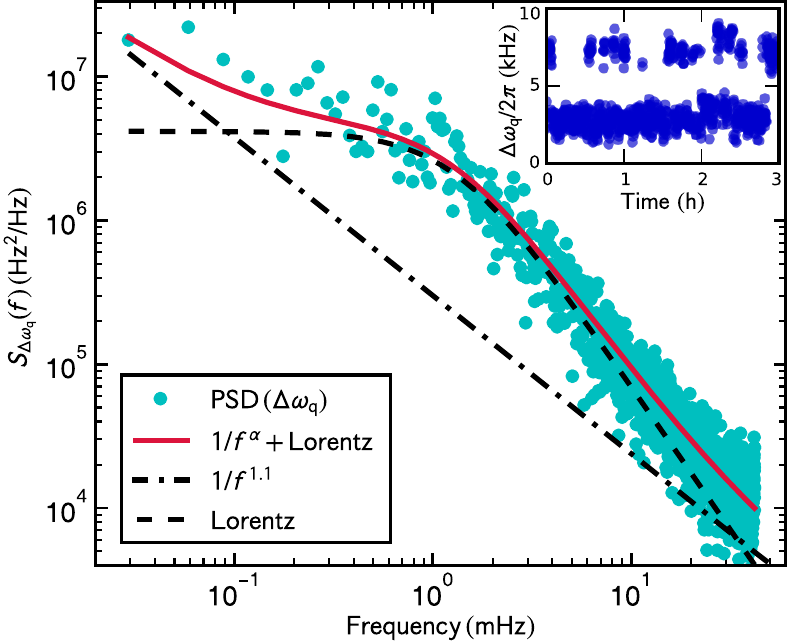}
		\protect\caption{\label{fig:Power-spectral-density_lorentz} Power spectral density of frequency fluctuations $\Delta \omega_{\mathrm{q}}$ (cyan dots) in a long-term measurement of $47\,\mathrm{h}$, revealing significant deviation from $1/f^\alpha$ noise (dash-dotted line). A fit (red solid line) is in agreement with the effect of a single thermal TLF (black dashed line) plus $1/f^{\alpha}$. The inset shows a short section of raw data, showing telegraphic noise that is presumably due to frequency switching of a near-resonant TLS coupled to a single thermal fluctuator. 
			The frequency uncertainty is approximately the size of the dots.
		} 
	\end{figure}

	The PSD of our measurements, shown in Fig.~\ref{fig:Power-spectral-density_lorentz}, deviates strongly from the ensemble $1/f$ noise limit, but is fit well by a single Lorentzian added to a $1/f^{\alpha}$-type background. From these measurements, we extract a background parameter of $\alpha\approx 1.1$ and the switching rate of the individual TLS of $\gamma_1\approx 1\,\mathrm{mHz}$. 
	For the distribution of switching rates, we estimate a TLF energy of $E_k/k_\mathrm{B}T=\ln(\Gamma_{\downarrow}/\Gamma_{\uparrow})=1.1$ in agreement with the assumption that the switching TLF are located spectrally close to the experimental temperature. For details on the PSD analysis, see appendix~D.

	Finally, we discuss the influence of other possible sources for discrete fluctuations: non-equilibrium quasiparticles (qp), movement of Abriskosov vortices and temperature fluctuations. The transmon qubit's transition energy is exponentially insensitive to charge fluctuations with respect to $\sqrt{E_{J}/E_{C}}$~\cite{Koch2007}. 
	In our sample, the change in qubit frequency due to a single qp, switching the charge parity of the capacitance~\cite{Catelani2014,Riste2013}, is about $2\,\mathrm{Hz}$ and thus not observable. A large number of non-equilibrium qp may contribute to relaxation~\cite{Gustavsson2016} but can not account for discrete fluctuations in $\omega_\mathrm{q}$ or abrupt changes in dynamics.  
	High magnetic fields may induce field dependent loss in a single junction qubit~\cite{Schneider2019}. To verify the intrinsic insensitivity of this experiment to flux noise, we measured the sample with roughly in-plane magnetic fields up to $\pm \SI{1}{\milli\tesla}$, and observed no changes in either coherence or frequency stability. Possible residual fields e.g.\ due to adsorbates~\cite{Kumar2016} are many orders of magnitude smaller. Significant correlation of the absolute fluctuation strength and the relaxation rate $[|\frac{\mathrm{d}}{\mathrm{d}t}\,\Delta \omega_{\mathrm{q}}| \star\,\Gamma_{1}]$ during periods of low dephasing times require transversal coupling, rendering direct influence of far detuned TLF and critical current fluctuations unlikely. Temperature fluctuations are known to induce low-frequency critical current noise~\cite{Anton2012a}. This effect is exponentially temperature dependent and found to be relevant at $T \gtrsim T_\mathrm{c}/3$ in $\mathrm{Al\text{-}AlO_{x}\text{-}Al}$ junctions. At our experimental temperature of $T = 20\,\mathrm{mK}$ its effect is several orders of magnitude below the observed noise level and can be excluded.

	In summary, we used a time-multiplexed protocol in long term measurements to extract correlated coherence information of a non-tunable transmon qubit. 
	We find positive and negative correlation between dephasing and fluctuations in qubit frequency on the timescale of seconds to days, which we attribute to the influence of individual dominant TLS, located close to conductor edges. Cross-correlation and PSD analysis confirm this interpretation and ascribe the source of fluctuation to interactions between thermal fluctuators and surface-TLS near resonance with the qubit. 
	
	Single defects reducing the coherence of qubits by up to one order of magnitude are a major challenge for future quantum computers. 
	Our findings make continuous re-calibration a necessity in today's solid-state qubits, although new materials or processing~\cite{Graaf2018,Bruno2015} might mitigate the problem. 
	However, our results imply that fundamental improvements of qubit parameter stability are necessary in order to realize useful many-qubit systems.
	
	$Note\,\,added-$During submission of this manuscript, a preprint on comparable 
	observations was published by Burnett \textit{et al.}~\cite{Burnett2019}, who independently  arrived at the conclusion that TLS are a major contribution to qubit parameter fluctuation.
	\\

	\begin{acknowledgments}
		We wish to thank M. Sandberg and M. Vissers for providing the sample  and D. Slichter and the group of J. Bylander for fruitful discussion. We gratefully acknowledge support by the ERC grant 648011, by DFG projects INST 121384/138-1 FUGG, WE 4359-7 and LI2446/1 (J.L.), the Swiss National Science Foundation through NCCR QSIT (C.M.), the Helmholtz International Research School for Teratronics (S.S.), the Carl-Zeiss-Foundation (A.S.), the National University of Science and Technology MISIS (Contract No. K2-2017-081), the NIST Quantum Information Initiative, ARO, IARPA, and DOE. This work is a contribution of the US government, and is not subject to copyright. 
		
	\end{acknowledgments}
	
%

	%\bibliography{C://Users//Steffen//Desktop//Thesis//bib//citations_Schloer}{}

\begin{thebibliography}{56}%
	\makeatletter
	\providecommand \@ifxundefined [1]{%
		\@ifx{#1\undefined}
	}%
	\providecommand \@ifnum [1]{%
		\ifnum #1\expandafter \@firstoftwo
		\else \expandafter \@secondoftwo
		\fi
	}%
	\providecommand \@ifx [1]{%
		\ifx #1\expandafter \@firstoftwo
		\else \expandafter \@secondoftwo
		\fi
	}%
	\providecommand \natexlab [1]{#1}%
	\providecommand \enquote  [1]{``#1''}%
	\providecommand \bibnamefont  [1]{#1}%
	\providecommand \bibfnamefont [1]{#1}%
	\providecommand \citenamefont [1]{#1}%
	\providecommand \href@noop [0]{\@secondoftwo}%
	\providecommand \href [0]{\begingroup \@sanitize@url \@href}%
	\providecommand \@href[1]{\@@startlink{#1}\@@href}%
	\providecommand \@@href[1]{\endgroup#1\@@endlink}%
	\providecommand \@sanitize@url [0]{\catcode `\\12\catcode `\$12\catcode
		`\&12\catcode `\#12\catcode `\^12\catcode `\_12\catcode `\%12\relax}%
	\providecommand \@@startlink[1]{}%
	\providecommand \@@endlink[0]{}%
	\providecommand \url  [0]{\begingroup\@sanitize@url \@url }%
	\providecommand \@url [1]{\endgroup\@href {#1}{\urlprefix }}%
	\providecommand \urlprefix  [0]{URL }%
	\providecommand \Eprint [0]{\href }%
	\providecommand \doibase [0]{http://dx.doi.org/}%
	\providecommand \selectlanguage [0]{\@gobble}%
	\providecommand \bibinfo  [0]{\@secondoftwo}%
	\providecommand \bibfield  [0]{\@secondoftwo}%
	\providecommand \translation [1]{[#1]}%
	\providecommand \BibitemOpen [0]{}%
	\providecommand \bibitemStop [0]{}%
	\providecommand \bibitemNoStop [0]{.\EOS\space}%
	\providecommand \EOS [0]{\spacefactor3000\relax}%
	\providecommand \BibitemShut  [1]{\csname bibitem#1\endcsname}%
	\let\auto@bib@innerbib\@empty
	%</preamble>
	\bibitem [{\citenamefont {Koch}\ \emph {et~al.}(2007)\citenamefont {Koch},
		\citenamefont {Yu}, \citenamefont {Gambetta}, \citenamefont {Houck},
		\citenamefont {Schuster}, \citenamefont {Majer}, \citenamefont {Blais},
		\citenamefont {Devoret}, \citenamefont {Girvin},\ and\ \citenamefont
		{Schoelkopf}}]{Koch2007}%
	\BibitemOpen
	\bibfield  {author} {\bibinfo {author} {\bibfnamefont {J.}~\bibnamefont
			{Koch}}, \bibinfo {author} {\bibfnamefont {T.~M.}\ \bibnamefont {Yu}},
		\bibinfo {author} {\bibfnamefont {J.}~\bibnamefont {Gambetta}}, \bibinfo
		{author} {\bibfnamefont {A.~A.}\ \bibnamefont {Houck}}, \bibinfo {author}
		{\bibfnamefont {D.~I.}\ \bibnamefont {Schuster}}, \bibinfo {author}
		{\bibfnamefont {J.}~\bibnamefont {Majer}}, \bibinfo {author} {\bibfnamefont
			{A.}~\bibnamefont {Blais}}, \bibinfo {author} {\bibfnamefont {M.~H.}\
			\bibnamefont {Devoret}}, \bibinfo {author} {\bibfnamefont {S.~M.}\
			\bibnamefont {Girvin}}, \ and\ \bibinfo {author} {\bibfnamefont {R.~J.}\
			\bibnamefont {Schoelkopf}},\ }\href {\doibase 10.1103/physreva.76.042319}
	{\bibfield  {journal} {\bibinfo  {journal} {Physical Review A}\ }\textbf
		{\bibinfo {volume} {76}} (\bibinfo {year} {2007}),\
		10.1103/physreva.76.042319}\BibitemShut {NoStop}%
	\bibitem [{\citenamefont {Kandala}\ \emph {et~al.}(2017)\citenamefont
		{Kandala}, \citenamefont {Mezzacapo}, \citenamefont {Temme}, \citenamefont
		{Takita}, \citenamefont {Brink}, \citenamefont {Chow},\ and\ \citenamefont
		{Gambetta}}]{Kandala2017}%
	\BibitemOpen
	\bibfield  {author} {\bibinfo {author} {\bibfnamefont {A.}~\bibnamefont
			{Kandala}}, \bibinfo {author} {\bibfnamefont {A.}~\bibnamefont {Mezzacapo}},
		\bibinfo {author} {\bibfnamefont {K.}~\bibnamefont {Temme}}, \bibinfo
		{author} {\bibfnamefont {M.}~\bibnamefont {Takita}}, \bibinfo {author}
		{\bibfnamefont {M.}~\bibnamefont {Brink}}, \bibinfo {author} {\bibfnamefont
			{J.~M.}\ \bibnamefont {Chow}}, \ and\ \bibinfo {author} {\bibfnamefont
			{J.~M.}\ \bibnamefont {Gambetta}},\ }\href {\doibase 10.1038/nature23879}
	{\bibfield  {journal} {\bibinfo  {journal} {Nature}\ }\textbf {\bibinfo
			{volume} {549}},\ \bibinfo {pages} {242} (\bibinfo {year}
		{2017})}\BibitemShut {NoStop}%
	\bibitem [{\citenamefont {Barends}\ \emph {et~al.}(2014)\citenamefont
		{Barends}, \citenamefont {Kelly}, \citenamefont {Megrant}, \citenamefont
		{Veitia}, \citenamefont {Sank}, \citenamefont {Jeffrey}, \citenamefont
		{White}, \citenamefont {Mutus}, \citenamefont {Fowler}, \citenamefont
		{Campbell}, \citenamefont {Chen}, \citenamefont {Chen}, \citenamefont
		{Chiaro}, \citenamefont {Dunsworth}, \citenamefont {Neill}, \citenamefont
		{O'Malley}, \citenamefont {Roushan}, \citenamefont {Vainsencher},
		\citenamefont {Wenner}, \citenamefont {Korotkov}, \citenamefont {Cleland},\
		and\ \citenamefont {Martinis}}]{Barends2014}%
	\BibitemOpen
	\bibfield  {author} {\bibinfo {author} {\bibfnamefont {R.}~\bibnamefont
			{Barends}}, \bibinfo {author} {\bibfnamefont {J.}~\bibnamefont {Kelly}},
		\bibinfo {author} {\bibfnamefont {A.}~\bibnamefont {Megrant}}, \bibinfo
		{author} {\bibfnamefont {A.}~\bibnamefont {Veitia}}, \bibinfo {author}
		{\bibfnamefont {D.}~\bibnamefont {Sank}}, \bibinfo {author} {\bibfnamefont
			{E.}~\bibnamefont {Jeffrey}}, \bibinfo {author} {\bibfnamefont {T.~C.}\
			\bibnamefont {White}}, \bibinfo {author} {\bibfnamefont {J.}~\bibnamefont
			{Mutus}}, \bibinfo {author} {\bibfnamefont {A.~G.}\ \bibnamefont {Fowler}},
		\bibinfo {author} {\bibfnamefont {B.}~\bibnamefont {Campbell}}, \bibinfo
		{author} {\bibfnamefont {Y.}~\bibnamefont {Chen}}, \bibinfo {author}
		{\bibfnamefont {Z.}~\bibnamefont {Chen}}, \bibinfo {author} {\bibfnamefont
			{B.}~\bibnamefont {Chiaro}}, \bibinfo {author} {\bibfnamefont
			{A.}~\bibnamefont {Dunsworth}}, \bibinfo {author} {\bibfnamefont
			{C.}~\bibnamefont {Neill}}, \bibinfo {author} {\bibfnamefont
			{P.}~\bibnamefont {O'Malley}}, \bibinfo {author} {\bibfnamefont
			{P.}~\bibnamefont {Roushan}}, \bibinfo {author} {\bibfnamefont
			{A.}~\bibnamefont {Vainsencher}}, \bibinfo {author} {\bibfnamefont
			{J.}~\bibnamefont {Wenner}}, \bibinfo {author} {\bibfnamefont {A.~N.}\
			\bibnamefont {Korotkov}}, \bibinfo {author} {\bibfnamefont {A.~N.}\
			\bibnamefont {Cleland}}, \ and\ \bibinfo {author} {\bibfnamefont {J.~M.}\
			\bibnamefont {Martinis}},\ }\href {\doibase 10.1038/nature13171} {\bibfield
		{journal} {\bibinfo  {journal} {Nature}\ }\textbf {\bibinfo {volume} {508}},\
		\bibinfo {pages} {500} (\bibinfo {year} {2014})}\BibitemShut {NoStop}%
	\bibitem [{\citenamefont {Fowler}\ \emph {et~al.}(2012)\citenamefont {Fowler},
		\citenamefont {Mariantoni}, \citenamefont {Martinis},\ and\ \citenamefont
		{Cleland}}]{Fowler2012}%
	\BibitemOpen
	\bibfield  {author} {\bibinfo {author} {\bibfnamefont {A.~G.}\ \bibnamefont
			{Fowler}}, \bibinfo {author} {\bibfnamefont {M.}~\bibnamefont {Mariantoni}},
		\bibinfo {author} {\bibfnamefont {J.~M.}\ \bibnamefont {Martinis}}, \ and\
		\bibinfo {author} {\bibfnamefont {A.~N.}\ \bibnamefont {Cleland}},\ }\href
	{\doibase 10.1103/physreva.86.032324} {\bibfield  {journal} {\bibinfo
			{journal} {Physical Review A}\ }\textbf {\bibinfo {volume} {86}} (\bibinfo
		{year} {2012}),\ 10.1103/physreva.86.032324}\BibitemShut {NoStop}%
	\bibitem [{\citenamefont {Klimov}\ \emph {et~al.}(2018)\citenamefont {Klimov},
		\citenamefont {Kelly}, \citenamefont {Chen}, \citenamefont {Neeley},
		\citenamefont {Megrant}, \citenamefont {Burkett}, \citenamefont {Barends},
		\citenamefont {Arya}, \citenamefont {Chiaro}, \citenamefont {Chen},
		\citenamefont {Dunsworth}, \citenamefont {Fowler}, \citenamefont {Foxen},
		\citenamefont {Gidney}, \citenamefont {Giustina}, \citenamefont {Graff},
		\citenamefont {Huang}, \citenamefont {Jeffrey}, \citenamefont {Lucero},
		\citenamefont {Mutus}, \citenamefont {Naaman}, \citenamefont {Neill},
		\citenamefont {Quintana}, \citenamefont {Roushan}, \citenamefont {Sank},
		\citenamefont {Vainsencher}, \citenamefont {Wenner}, \citenamefont {White},
		\citenamefont {Boixo}, \citenamefont {Babbush}, \citenamefont {Smelyanskiy},
		\citenamefont {Neven},\ and\ \citenamefont {Martinis}}]{Klimov2018}%
	\BibitemOpen
	\bibfield  {author} {\bibinfo {author} {\bibfnamefont {P.}~\bibnamefont
			{Klimov}}, \bibinfo {author} {\bibfnamefont {J.}~\bibnamefont {Kelly}},
		\bibinfo {author} {\bibfnamefont {Z.}~\bibnamefont {Chen}}, \bibinfo {author}
		{\bibfnamefont {M.}~\bibnamefont {Neeley}}, \bibinfo {author} {\bibfnamefont
			{A.}~\bibnamefont {Megrant}}, \bibinfo {author} {\bibfnamefont
			{B.}~\bibnamefont {Burkett}}, \bibinfo {author} {\bibfnamefont
			{R.}~\bibnamefont {Barends}}, \bibinfo {author} {\bibfnamefont
			{K.}~\bibnamefont {Arya}}, \bibinfo {author} {\bibfnamefont {B.}~\bibnamefont
			{Chiaro}}, \bibinfo {author} {\bibfnamefont {Y.}~\bibnamefont {Chen}},
		\bibinfo {author} {\bibfnamefont {A.}~\bibnamefont {Dunsworth}}, \bibinfo
		{author} {\bibfnamefont {A.}~\bibnamefont {Fowler}}, \bibinfo {author}
		{\bibfnamefont {B.}~\bibnamefont {Foxen}}, \bibinfo {author} {\bibfnamefont
			{C.}~\bibnamefont {Gidney}}, \bibinfo {author} {\bibfnamefont
			{M.}~\bibnamefont {Giustina}}, \bibinfo {author} {\bibfnamefont
			{R.}~\bibnamefont {Graff}}, \bibinfo {author} {\bibfnamefont
			{T.}~\bibnamefont {Huang}}, \bibinfo {author} {\bibfnamefont
			{E.}~\bibnamefont {Jeffrey}}, \bibinfo {author} {\bibfnamefont
			{E.}~\bibnamefont {Lucero}}, \bibinfo {author} {\bibfnamefont
			{J.}~\bibnamefont {Mutus}}, \bibinfo {author} {\bibfnamefont
			{O.}~\bibnamefont {Naaman}}, \bibinfo {author} {\bibfnamefont
			{C.}~\bibnamefont {Neill}}, \bibinfo {author} {\bibfnamefont
			{C.}~\bibnamefont {Quintana}}, \bibinfo {author} {\bibfnamefont
			{P.}~\bibnamefont {Roushan}}, \bibinfo {author} {\bibfnamefont
			{D.}~\bibnamefont {Sank}}, \bibinfo {author} {\bibfnamefont {A.}~\bibnamefont
			{Vainsencher}}, \bibinfo {author} {\bibfnamefont {J.}~\bibnamefont {Wenner}},
		\bibinfo {author} {\bibfnamefont {T.}~\bibnamefont {White}}, \bibinfo
		{author} {\bibfnamefont {S.}~\bibnamefont {Boixo}}, \bibinfo {author}
		{\bibfnamefont {R.}~\bibnamefont {Babbush}}, \bibinfo {author} {\bibfnamefont
			{V.}~\bibnamefont {Smelyanskiy}}, \bibinfo {author} {\bibfnamefont
			{H.}~\bibnamefont {Neven}}, \ and\ \bibinfo {author} {\bibfnamefont
			{J.}~\bibnamefont {Martinis}},\ }\href {\doibase
		10.1103/physrevlett.121.090502} {\bibfield  {journal} {\bibinfo  {journal}
			{Physical Review Letters}\ }\textbf {\bibinfo {volume} {121}} (\bibinfo
		{year} {2018}),\ 10.1103/physrevlett.121.090502}\BibitemShut {NoStop}%
	\bibitem [{\citenamefont {Paik}\ \emph {et~al.}(2011)\citenamefont {Paik},
		\citenamefont {Schuster}, \citenamefont {Bishop}, \citenamefont {Kirchmair},
		\citenamefont {Catelani}, \citenamefont {Sears}, \citenamefont {Johnson},
		\citenamefont {Reagor}, \citenamefont {Frunzio}, \citenamefont {Glazman},
		\citenamefont {Girvin}, \citenamefont {Devoret},\ and\ \citenamefont
		{Schoelkopf}}]{Paik2011}%
	\BibitemOpen
	\bibfield  {author} {\bibinfo {author} {\bibfnamefont {H.}~\bibnamefont
			{Paik}}, \bibinfo {author} {\bibfnamefont {D.~I.}\ \bibnamefont {Schuster}},
		\bibinfo {author} {\bibfnamefont {L.~S.}\ \bibnamefont {Bishop}}, \bibinfo
		{author} {\bibfnamefont {G.}~\bibnamefont {Kirchmair}}, \bibinfo {author}
		{\bibfnamefont {G.}~\bibnamefont {Catelani}}, \bibinfo {author}
		{\bibfnamefont {A.~P.}\ \bibnamefont {Sears}}, \bibinfo {author}
		{\bibfnamefont {B.~R.}\ \bibnamefont {Johnson}}, \bibinfo {author}
		{\bibfnamefont {M.~J.}\ \bibnamefont {Reagor}}, \bibinfo {author}
		{\bibfnamefont {L.}~\bibnamefont {Frunzio}}, \bibinfo {author} {\bibfnamefont
			{L.~I.}\ \bibnamefont {Glazman}}, \bibinfo {author} {\bibfnamefont {S.~M.}\
			\bibnamefont {Girvin}}, \bibinfo {author} {\bibfnamefont {M.~H.}\
			\bibnamefont {Devoret}}, \ and\ \bibinfo {author} {\bibfnamefont {R.~J.}\
			\bibnamefont {Schoelkopf}},\ }\href {\doibase 10.1103/physrevlett.107.240501}
	{\bibfield  {journal} {\bibinfo  {journal} {Physical Review Letters}\
		}\textbf {\bibinfo {volume} {107}} (\bibinfo {year} {2011}),\
		10.1103/physrevlett.107.240501}\BibitemShut {NoStop}%
	\bibitem [{\citenamefont {Dial}\ \emph {et~al.}(2016)\citenamefont {Dial},
		\citenamefont {McClure}, \citenamefont {Poletto}, \citenamefont {Keefe},
		\citenamefont {Rothwell}, \citenamefont {Gambetta}, \citenamefont {Abraham},
		\citenamefont {Chow},\ and\ \citenamefont {Steffen}}]{Dial2016}%
	\BibitemOpen
	\bibfield  {author} {\bibinfo {author} {\bibfnamefont {O.}~\bibnamefont
			{Dial}}, \bibinfo {author} {\bibfnamefont {D.~T.}\ \bibnamefont {McClure}},
		\bibinfo {author} {\bibfnamefont {S.}~\bibnamefont {Poletto}}, \bibinfo
		{author} {\bibfnamefont {G.~A.}\ \bibnamefont {Keefe}}, \bibinfo {author}
		{\bibfnamefont {M.~B.}\ \bibnamefont {Rothwell}}, \bibinfo {author}
		{\bibfnamefont {J.~M.}\ \bibnamefont {Gambetta}}, \bibinfo {author}
		{\bibfnamefont {D.~W.}\ \bibnamefont {Abraham}}, \bibinfo {author}
		{\bibfnamefont {J.~M.}\ \bibnamefont {Chow}}, \ and\ \bibinfo {author}
		{\bibfnamefont {M.}~\bibnamefont {Steffen}},\ }\href@noop {} {\bibfield
		{journal} {\bibinfo  {journal} {Superconductor Science and Technology}\
		}\textbf {\bibinfo {volume} {29}},\ \bibinfo {pages} {044001} (\bibinfo
		{year} {2016})}\BibitemShut {NoStop}%
	\bibitem [{\citenamefont {Anderson}, \citenamefont {Halperin},\ and\
		\citenamefont {Varma}(1972)}]{Anderson1972}%
	\BibitemOpen
	\bibfield  {author} {\bibinfo {author} {\bibfnamefont {P.~W.}\ \bibnamefont
			{Anderson}}, \bibinfo {author} {\bibfnamefont {B.~I.}\ \bibnamefont
			{Halperin}}, \ and\ \bibinfo {author} {\bibfnamefont {C.~M.}\ \bibnamefont
			{Varma}},\ }\href {\doibase 10.1080/14786437208229210} {\bibfield  {journal}
		{\bibinfo  {journal} {Philosophical Magazine}\ }\textbf {\bibinfo {volume}
			{25}},\ \bibinfo {pages} {1} (\bibinfo {year} {1972})}\BibitemShut {NoStop}%
	\bibitem [{\citenamefont {Phillips}(1987)}]{Phillips:1987}%
	\BibitemOpen
	\bibfield  {author} {\bibinfo {author} {\bibfnamefont {W.~A.}\ \bibnamefont
			{Phillips}},\ }\href {\doibase 10.1088/0034-4885/50/12/003} {\bibfield
		{journal} {\bibinfo  {journal} {Reports on Progress in Physics}\ }\textbf
		{\bibinfo {volume} {50}},\ \bibinfo {pages} {1657} (\bibinfo {year}
		{1987})}\BibitemShut {NoStop}%
	\bibitem [{\citenamefont {M{\"u}ller}, \citenamefont {Cole},\ and\
		\citenamefont {Lisenfeld}(2017)}]{Mueller2017}%
	\BibitemOpen
	\bibfield  {author} {\bibinfo {author} {\bibfnamefont {C.}~\bibnamefont
			{M{\"u}ller}}, \bibinfo {author} {\bibfnamefont {J.~H.}\ \bibnamefont
			{Cole}}, \ and\ \bibinfo {author} {\bibfnamefont {J.}~\bibnamefont
			{Lisenfeld}},\ }\href {http://arxiv.org/abs/1705.01108v2} {\bibfield
		{journal} {\bibinfo  {journal} {arXiv}\ } (\bibinfo {year}
		{2017})}\BibitemShut {NoStop}%
	\bibitem [{\citenamefont {{Holder, Aaron M. and Osborn, Kevin D. and Lobb, C.
				J. and Musgrave, Charles B.}}(2013)}]{Holder:PRL:2013}%
	\BibitemOpen
	\bibfield  {author} {\bibinfo {author} {\bibnamefont {{Holder, Aaron M. and
					Osborn, Kevin D. and Lobb, C. J. and Musgrave, Charles B.}}},\ }\href@noop {}
	{\bibfield  {journal} {\bibinfo  {journal} {Physical Review Letters}\
		}\textbf {\bibinfo {volume} {111}},\ \bibinfo {pages} {065901} (\bibinfo
		{year} {2013})}\BibitemShut {NoStop}%
	\bibitem [{\citenamefont {de~Graaf}\ \emph {et~al.}(2018)\citenamefont
		{de~Graaf}, \citenamefont {Faoro}, \citenamefont {Burnett}, \citenamefont
		{Adamyan}, \citenamefont {Tzalenchuk}, \citenamefont {Kubatkin},
		\citenamefont {Lindstr{\"o}m},\ and\ \citenamefont {Danilov}}]{Graaf2018}%
	\BibitemOpen
	\bibfield  {author} {\bibinfo {author} {\bibfnamefont {S.~E.}\ \bibnamefont
			{de~Graaf}}, \bibinfo {author} {\bibfnamefont {L.}~\bibnamefont {Faoro}},
		\bibinfo {author} {\bibfnamefont {J.}~\bibnamefont {Burnett}}, \bibinfo
		{author} {\bibfnamefont {A.~A.}\ \bibnamefont {Adamyan}}, \bibinfo {author}
		{\bibfnamefont {A.~Y.}\ \bibnamefont {Tzalenchuk}}, \bibinfo {author}
		{\bibfnamefont {S.~E.}\ \bibnamefont {Kubatkin}}, \bibinfo {author}
		{\bibfnamefont {T.}~\bibnamefont {Lindstr{\"o}m}}, \ and\ \bibinfo {author}
		{\bibfnamefont {A.~V.}\ \bibnamefont {Danilov}},\ }\href {\doibase
		10.1038/s41467-018-03577-2} {\bibfield  {journal} {\bibinfo  {journal}
			{Nature Communications}\ }\textbf {\bibinfo {volume} {9}} (\bibinfo {year}
		{2018}),\ 10.1038/s41467-018-03577-2}\BibitemShut {NoStop}%
	\bibitem [{\citenamefont {Faoro}\ and\ \citenamefont
		{Ioffe}(2015)}]{Faoro2015}%
	\BibitemOpen
	\bibfield  {author} {\bibinfo {author} {\bibfnamefont {L.}~\bibnamefont
			{Faoro}}\ and\ \bibinfo {author} {\bibfnamefont {L.~B.}\ \bibnamefont
			{Ioffe}},\ }\href {\doibase 10.1103/physrevb.91.014201} {\bibfield  {journal}
		{\bibinfo  {journal} {Physical Review B}\ }\textbf {\bibinfo {volume} {91}}
		(\bibinfo {year} {2015}),\ 10.1103/physrevb.91.014201}\BibitemShut {NoStop}%
	\bibitem [{\citenamefont {Martinis}\ \emph {et~al.}(2005)\citenamefont
		{Martinis}, \citenamefont {Cooper}, \citenamefont {McDermott}, \citenamefont
		{Steffen}, \citenamefont {Ansmann}, \citenamefont {Osborn}, \citenamefont
		{Cicak}, \citenamefont {Oh}, \citenamefont {Pappas}, \citenamefont
		{Simmonds},\ and\ \citenamefont {Yu}}]{Martinis2005}%
	\BibitemOpen
	\bibfield  {author} {\bibinfo {author} {\bibfnamefont {J.~M.}\ \bibnamefont
			{Martinis}}, \bibinfo {author} {\bibfnamefont {K.~B.}\ \bibnamefont
			{Cooper}}, \bibinfo {author} {\bibfnamefont {R.}~\bibnamefont {McDermott}},
		\bibinfo {author} {\bibfnamefont {M.}~\bibnamefont {Steffen}}, \bibinfo
		{author} {\bibfnamefont {M.}~\bibnamefont {Ansmann}}, \bibinfo {author}
		{\bibfnamefont {K.~D.}\ \bibnamefont {Osborn}}, \bibinfo {author}
		{\bibfnamefont {K.}~\bibnamefont {Cicak}}, \bibinfo {author} {\bibfnamefont
			{S.}~\bibnamefont {Oh}}, \bibinfo {author} {\bibfnamefont {D.~P.}\
			\bibnamefont {Pappas}}, \bibinfo {author} {\bibfnamefont {R.~W.}\
			\bibnamefont {Simmonds}}, \ and\ \bibinfo {author} {\bibfnamefont {C.~C.}\
			\bibnamefont {Yu}},\ }\href {\doibase 10.1103/physrevlett.95.210503}
	{\bibfield  {journal} {\bibinfo  {journal} {Physical Review Letters}\
		}\textbf {\bibinfo {volume} {95}} (\bibinfo {year} {2005}),\
		10.1103/physrevlett.95.210503}\BibitemShut {NoStop}%
	\bibitem [{\citenamefont {Weides}\ \emph {et~al.}(2011)\citenamefont {Weides},
		\citenamefont {Kline}, \citenamefont {Vissers}, \citenamefont {Sandberg},
		\citenamefont {Wisbey}, \citenamefont {Johnson}, \citenamefont {Ohki},\ and\
		\citenamefont {Pappas}}]{Weides2011}%
	\BibitemOpen
	\bibfield  {author} {\bibinfo {author} {\bibfnamefont {M.~P.}\ \bibnamefont
			{Weides}}, \bibinfo {author} {\bibfnamefont {J.~S.}\ \bibnamefont {Kline}},
		\bibinfo {author} {\bibfnamefont {M.~R.}\ \bibnamefont {Vissers}}, \bibinfo
		{author} {\bibfnamefont {M.~O.}\ \bibnamefont {Sandberg}}, \bibinfo {author}
		{\bibfnamefont {D.~S.}\ \bibnamefont {Wisbey}}, \bibinfo {author}
		{\bibfnamefont {B.~R.}\ \bibnamefont {Johnson}}, \bibinfo {author}
		{\bibfnamefont {T.~A.}\ \bibnamefont {Ohki}}, \ and\ \bibinfo {author}
		{\bibfnamefont {D.~P.}\ \bibnamefont {Pappas}},\ }\href {\doibase
		10.1063/1.3672000} {\bibfield  {journal} {\bibinfo  {journal} {Applied
				Physics Letters}\ }\textbf {\bibinfo {volume} {99}},\ \bibinfo {pages}
		{262502} (\bibinfo {year} {2011})}\BibitemShut {NoStop}%
	\bibitem [{\citenamefont {{M. Weides and R. C. Bialczak and M. Lenander and E.
				Lucero and Matteo Mariantoni and M. Neeley and A. D. O'Connell and D. Sank
				and H. Wang and J. Wenner and T. Yamamoto and Y. Yin and A. N. Cleland and J.
				Martinis}}(2011)}]{Weides2011a}%
	\BibitemOpen
	\bibfield  {author} {\bibinfo {author} {\bibnamefont {{M. Weides and R. C.
					Bialczak and M. Lenander and E. Lucero and Matteo Mariantoni and M. Neeley
					and A. D. O'Connell and D. Sank and H. Wang and J. Wenner and T. Yamamoto and
					Y. Yin and A. N. Cleland and J. Martinis}}},\ }\href {\doibase
		10.1088/0953-2048/24/5/055005} {\bibfield  {journal} {\bibinfo  {journal}
			{Superconductor Science and Technology}\ }\textbf {\bibinfo {volume} {24}},\
		\bibinfo {pages} {055005} (\bibinfo {year} {2011})}\BibitemShut {NoStop}%
	\bibitem [{\citenamefont {Dutta}\ and\ \citenamefont {Horn}(1981)}]{Dutta1981}%
	\BibitemOpen
	\bibfield  {author} {\bibinfo {author} {\bibfnamefont {P.}~\bibnamefont
			{Dutta}}\ and\ \bibinfo {author} {\bibfnamefont {P.~M.}\ \bibnamefont
			{Horn}},\ }\href {\doibase 10.1103/revmodphys.53.497} {\bibfield  {journal}
		{\bibinfo  {journal} {Reviews of Modern Physics}\ }\textbf {\bibinfo {volume}
			{53}},\ \bibinfo {pages} {497} (\bibinfo {year} {1981})}\BibitemShut
	{NoStop}%
	\bibitem [{\citenamefont {M{\"u}ller}\ \emph {et~al.}(2015)\citenamefont
		{M{\"u}ller}, \citenamefont {Lisenfeld}, \citenamefont {Shnirman},\ and\
		\citenamefont {Poletto}}]{Mueller2015}%
	\BibitemOpen
	\bibfield  {author} {\bibinfo {author} {\bibfnamefont {C.}~\bibnamefont
			{M{\"u}ller}}, \bibinfo {author} {\bibfnamefont {J.}~\bibnamefont
			{Lisenfeld}}, \bibinfo {author} {\bibfnamefont {A.}~\bibnamefont {Shnirman}},
		\ and\ \bibinfo {author} {\bibfnamefont {S.}~\bibnamefont {Poletto}},\ }\href
	{\doibase 10.1103/physrevb.92.035442} {\bibfield  {journal} {\bibinfo
			{journal} {Physical Review B}\ }\textbf {\bibinfo {volume} {92}} (\bibinfo
		{year} {2015}),\ 10.1103/physrevb.92.035442}\BibitemShut {NoStop}%
	\bibitem [{\citenamefont {Mei{\ss}ner}\ \emph {et~al.}(2018)\citenamefont
		{Mei{\ss}ner}, \citenamefont {Seiler}, \citenamefont {Lisenfeld},
		\citenamefont {Ustinov},\ and\ \citenamefont {Weiss}}]{Meisner2018}%
	\BibitemOpen
	\bibfield  {author} {\bibinfo {author} {\bibfnamefont {S.~M.}\ \bibnamefont
			{Mei{\ss}ner}}, \bibinfo {author} {\bibfnamefont {A.}~\bibnamefont {Seiler}},
		\bibinfo {author} {\bibfnamefont {J.}~\bibnamefont {Lisenfeld}}, \bibinfo
		{author} {\bibfnamefont {A.~V.}\ \bibnamefont {Ustinov}}, \ and\ \bibinfo
		{author} {\bibfnamefont {G.}~\bibnamefont {Weiss}},\ }\href {\doibase
		10.1103/physrevb.97.180505} {\bibfield  {journal} {\bibinfo  {journal}
			{Physical Review B}\ }\textbf {\bibinfo {volume} {97}} (\bibinfo {year}
		{2018}),\ 10.1103/physrevb.97.180505}\BibitemShut {NoStop}%
	\bibitem [{\citenamefont {Grabovskij}\ \emph {et~al.}(2012)\citenamefont
		{Grabovskij}, \citenamefont {Peichl}, \citenamefont {Lisenfeld},
		\citenamefont {Weiss},\ and\ \citenamefont {Ustinov}}]{Grabovskij2012}%
	\BibitemOpen
	\bibfield  {author} {\bibinfo {author} {\bibfnamefont {G.~J.}\ \bibnamefont
			{Grabovskij}}, \bibinfo {author} {\bibfnamefont {T.}~\bibnamefont {Peichl}},
		\bibinfo {author} {\bibfnamefont {J.}~\bibnamefont {Lisenfeld}}, \bibinfo
		{author} {\bibfnamefont {G.}~\bibnamefont {Weiss}}, \ and\ \bibinfo {author}
		{\bibfnamefont {A.~V.}\ \bibnamefont {Ustinov}},\ }\href {\doibase
		10.1126/science.1226487} {\bibfield  {journal} {\bibinfo  {journal}
			{Science}\ }\textbf {\bibinfo {volume} {338}},\ \bibinfo {pages} {232}
		(\bibinfo {year} {2012})}\BibitemShut {NoStop}%
	\bibitem [{\citenamefont {Lisenfeld}\ \emph {et~al.}(2015)\citenamefont
		{Lisenfeld}, \citenamefont {Grabovskij}, \citenamefont {M{\"u}ller},
		\citenamefont {Cole}, \citenamefont {Weiss},\ and\ \citenamefont
		{Ustinov}}]{Lisenfeld2015}%
	\BibitemOpen
	\bibfield  {author} {\bibinfo {author} {\bibfnamefont {J.}~\bibnamefont
			{Lisenfeld}}, \bibinfo {author} {\bibfnamefont {G.~J.}\ \bibnamefont
			{Grabovskij}}, \bibinfo {author} {\bibfnamefont {C.}~\bibnamefont
			{M{\"u}ller}}, \bibinfo {author} {\bibfnamefont {J.~H.}\ \bibnamefont
			{Cole}}, \bibinfo {author} {\bibfnamefont {G.}~\bibnamefont {Weiss}}, \ and\
		\bibinfo {author} {\bibfnamefont {A.~V.}\ \bibnamefont {Ustinov}},\ }\href
	{\doibase 10.1038/ncomms7182} {\bibfield  {journal} {\bibinfo  {journal}
			{Nature Communications}\ }\textbf {\bibinfo {volume} {6}},\ \bibinfo {pages}
		{6182} (\bibinfo {year} {2015})}\BibitemShut {NoStop}%
	\bibitem [{\citenamefont {Black}\ and\ \citenamefont
		{Halperin}(1977)}]{Black:PRB:1977}%
	\BibitemOpen
	\bibfield  {author} {\bibinfo {author} {\bibfnamefont {J.~L.}\ \bibnamefont
			{Black}}\ and\ \bibinfo {author} {\bibfnamefont {B.~I.}\ \bibnamefont
			{Halperin}},\ }\href@noop {} {\bibfield  {journal} {\bibinfo  {journal}
			{Physical Review B}\ }\textbf {\bibinfo {volume} {16}},\ \bibinfo {pages}
		{2879} (\bibinfo {year} {1977})}\BibitemShut {NoStop}%
	\bibitem [{\citenamefont {Faoro}\ and\ \citenamefont
		{Ioffe}(2012)}]{Faoro:PRL:2012}%
	\BibitemOpen
	\bibfield  {author} {\bibinfo {author} {\bibfnamefont {L.}~\bibnamefont
			{Faoro}}\ and\ \bibinfo {author} {\bibfnamefont {L.~B.}\ \bibnamefont
			{Ioffe}},\ }\href {\doibase 10.1103/PhysRevLett.109.157005} {\bibfield
		{journal} {\bibinfo  {journal} {Physical Review Letters}\ }\textbf {\bibinfo
			{volume} {109}},\ \bibinfo {pages} {157005} (\bibinfo {year}
		{2012})}\BibitemShut {NoStop}%
	\bibitem [{\citenamefont {Shnirman}\ \emph {et~al.}(2005)\citenamefont
		{Shnirman}, \citenamefont {Sch{\"o}n}, \citenamefont {Martin},\ and\
		\citenamefont {Makhlin}}]{Shnirman2005}%
	\BibitemOpen
	\bibfield  {author} {\bibinfo {author} {\bibfnamefont {A.}~\bibnamefont
			{Shnirman}}, \bibinfo {author} {\bibfnamefont {G.}~\bibnamefont {Sch{\"o}n}},
		\bibinfo {author} {\bibfnamefont {I.}~\bibnamefont {Martin}}, \ and\ \bibinfo
		{author} {\bibfnamefont {Y.}~\bibnamefont {Makhlin}},\ }\href {\doibase
		10.1103/physrevlett.94.127002} {\bibfield  {journal} {\bibinfo  {journal}
			{Physical Review Letters}\ }\textbf {\bibinfo {volume} {94}} (\bibinfo {year}
		{2005}),\ 10.1103/physrevlett.94.127002}\BibitemShut {NoStop}%
	\bibitem [{\citenamefont {Cole}\ \emph {et~al.}(2010)\citenamefont {Cole},
		\citenamefont {Müller}, \citenamefont {Bushev}, \citenamefont {Grabovskij},
		\citenamefont {Lisenfeld}, \citenamefont {Lukashenko}, \citenamefont
		{Ustinov},\ and\ \citenamefont {Shnirman}}]{Cole2010}%
	\BibitemOpen
	\bibfield  {author} {\bibinfo {author} {\bibfnamefont {J.~H.}\ \bibnamefont
			{Cole}}, \bibinfo {author} {\bibfnamefont {C.}~\bibnamefont {Müller}},
		\bibinfo {author} {\bibfnamefont {P.}~\bibnamefont {Bushev}}, \bibinfo
		{author} {\bibfnamefont {G.~J.}\ \bibnamefont {Grabovskij}}, \bibinfo
		{author} {\bibfnamefont {J.}~\bibnamefont {Lisenfeld}}, \bibinfo {author}
		{\bibfnamefont {A.}~\bibnamefont {Lukashenko}}, \bibinfo {author}
		{\bibfnamefont {A.~V.}\ \bibnamefont {Ustinov}}, \ and\ \bibinfo {author}
		{\bibfnamefont {A.}~\bibnamefont {Shnirman}},\ }\href {\doibase
		10.1063/1.3529457} {\bibfield  {journal} {\bibinfo  {journal} {Applied
				Physics Letters}\ }\textbf {\bibinfo {volume} {97}},\ \bibinfo {pages}
		{252501} (\bibinfo {year} {2010})}\BibitemShut {NoStop}%
	\bibitem [{\citenamefont {Sarabi}\ \emph {et~al.}(2016)\citenamefont {Sarabi},
		\citenamefont {Ramanayaka}, \citenamefont {Burin}, \citenamefont
		{Wellstood},\ and\ \citenamefont {Osborn}}]{Sarabi2016}%
	\BibitemOpen
	\bibfield  {author} {\bibinfo {author} {\bibfnamefont {B.}~\bibnamefont
			{Sarabi}}, \bibinfo {author} {\bibfnamefont {A.}~\bibnamefont {Ramanayaka}},
		\bibinfo {author} {\bibfnamefont {A.}~\bibnamefont {Burin}}, \bibinfo
		{author} {\bibfnamefont {F.}~\bibnamefont {Wellstood}}, \ and\ \bibinfo
		{author} {\bibfnamefont {K.}~\bibnamefont {Osborn}},\ }\href {\doibase
		10.1103/physrevlett.116.167002} {\bibfield  {journal} {\bibinfo  {journal}
			{Physical Review Letters}\ }\textbf {\bibinfo {volume} {116}} (\bibinfo
		{year} {2016}),\ 10.1103/physrevlett.116.167002}\BibitemShut {NoStop}%
	\bibitem [{\citenamefont {Brehm}\ \emph {et~al.}(2017)\citenamefont {Brehm},
		\citenamefont {Bilmes}, \citenamefont {Weiss}, \citenamefont {Ustinov},\ and\
		\citenamefont {Lisenfeld}}]{Brehm2017}%
	\BibitemOpen
	\bibfield  {author} {\bibinfo {author} {\bibfnamefont {J.~D.}\ \bibnamefont
			{Brehm}}, \bibinfo {author} {\bibfnamefont {A.}~\bibnamefont {Bilmes}},
		\bibinfo {author} {\bibfnamefont {G.}~\bibnamefont {Weiss}}, \bibinfo
		{author} {\bibfnamefont {A.~V.}\ \bibnamefont {Ustinov}}, \ and\ \bibinfo
		{author} {\bibfnamefont {J.}~\bibnamefont {Lisenfeld}},\ }\href {\doibase
		10.1063/1.5001920} {\bibfield  {journal} {\bibinfo  {journal} {Applied
				Physics Letters}\ }\textbf {\bibinfo {volume} {111}},\ \bibinfo {pages}
		{112601} (\bibinfo {year} {2017})}\BibitemShut {NoStop}%
	\bibitem [{\citenamefont {Lisenfeld}\ \emph {et~al.}(2016)\citenamefont
		{Lisenfeld}, \citenamefont {Bilmes}, \citenamefont {Matityahu}, \citenamefont
		{Zanker}, \citenamefont {Marthaler}, \citenamefont {Schechter}, \citenamefont
		{Sch{\"o}n}, \citenamefont {Shnirman}, \citenamefont {Weiss},\ and\
		\citenamefont {Ustinov}}]{Lisenfeld2016}%
	\BibitemOpen
	\bibfield  {author} {\bibinfo {author} {\bibfnamefont {J.}~\bibnamefont
			{Lisenfeld}}, \bibinfo {author} {\bibfnamefont {A.}~\bibnamefont {Bilmes}},
		\bibinfo {author} {\bibfnamefont {S.}~\bibnamefont {Matityahu}}, \bibinfo
		{author} {\bibfnamefont {S.}~\bibnamefont {Zanker}}, \bibinfo {author}
		{\bibfnamefont {M.}~\bibnamefont {Marthaler}}, \bibinfo {author}
		{\bibfnamefont {M.}~\bibnamefont {Schechter}}, \bibinfo {author}
		{\bibfnamefont {G.}~\bibnamefont {Sch{\"o}n}}, \bibinfo {author}
		{\bibfnamefont {A.}~\bibnamefont {Shnirman}}, \bibinfo {author}
		{\bibfnamefont {G.}~\bibnamefont {Weiss}}, \ and\ \bibinfo {author}
		{\bibfnamefont {A.~V.}\ \bibnamefont {Ustinov}},\ }\href {\doibase
		10.1038/srep23786} {\bibfield  {journal} {\bibinfo  {journal} {Scientific
				Reports}\ }\textbf {\bibinfo {volume} {6}} (\bibinfo {year} {2016}),\
		10.1038/srep23786}\BibitemShut {NoStop}%
	\bibitem [{\citenamefont {{D. H. Slichter and C. Müller and R. Vijay and S. J.
				Weber and A. Blais and I. Siddiqi}}(2016)}]{Slichter2016}%
	\BibitemOpen
	\bibfield  {author} {\bibinfo {author} {\bibnamefont {{D. H. Slichter and C.
					Müller and R. Vijay and S. J. Weber and A. Blais and I. Siddiqi}}},\ }\href
	{\doibase 10.1088/1367-2630/18/5/053031} {\bibfield  {journal} {\bibinfo
			{journal} {New Journal of Physics}\ }\textbf {\bibinfo {volume} {18}},\
		\bibinfo {pages} {053031} (\bibinfo {year} {2016})}\BibitemShut {NoStop}%
	\bibitem [{\citenamefont {Asban}\ \emph {et~al.}(2017)\citenamefont {Asban},
		\citenamefont {Amir}, \citenamefont {Imry},\ and\ \citenamefont
		{Schechter}}]{Asban2017}%
	\BibitemOpen
	\bibfield  {author} {\bibinfo {author} {\bibfnamefont {O.}~\bibnamefont
			{Asban}}, \bibinfo {author} {\bibfnamefont {A.}~\bibnamefont {Amir}},
		\bibinfo {author} {\bibfnamefont {Y.}~\bibnamefont {Imry}}, \ and\ \bibinfo
		{author} {\bibfnamefont {M.}~\bibnamefont {Schechter}},\ }\href {\doibase
		10.1103/physrevb.95.144207} {\bibfield  {journal} {\bibinfo  {journal}
			{Physical Review B}\ }\textbf {\bibinfo {volume} {95}} (\bibinfo {year}
		{2017}),\ 10.1103/physrevb.95.144207}\BibitemShut {NoStop}%
	\bibitem [{\citenamefont {Catelani}(2014)}]{Catelani2014}%
	\BibitemOpen
	\bibfield  {author} {\bibinfo {author} {\bibfnamefont {G.}~\bibnamefont
			{Catelani}},\ }\href {\doibase 10.1103/physrevb.89.094522} {\bibfield
		{journal} {\bibinfo  {journal} {Physical Review B}\ }\textbf {\bibinfo
			{volume} {89}} (\bibinfo {year} {2014}),\
		10.1103/physrevb.89.094522}\BibitemShut {NoStop}%
	\bibitem [{\citenamefont {Rist{\`{e}}}\ \emph {et~al.}(2013)\citenamefont
		{Rist{\`{e}}}, \citenamefont {Bultink}, \citenamefont {Tiggelman},
		\citenamefont {Schouten}, \citenamefont {Lehnert},\ and\ \citenamefont
		{DiCarlo}}]{Riste2013}%
	\BibitemOpen
	\bibfield  {author} {\bibinfo {author} {\bibfnamefont {D.}~\bibnamefont
			{Rist{\`{e}}}}, \bibinfo {author} {\bibfnamefont {C.~C.}\ \bibnamefont
			{Bultink}}, \bibinfo {author} {\bibfnamefont {M.~J.}\ \bibnamefont
			{Tiggelman}}, \bibinfo {author} {\bibfnamefont {R.~N.}\ \bibnamefont
			{Schouten}}, \bibinfo {author} {\bibfnamefont {K.~W.}\ \bibnamefont
			{Lehnert}}, \ and\ \bibinfo {author} {\bibfnamefont {L.}~\bibnamefont
			{DiCarlo}},\ }\href {\doibase 10.1038/ncomms2936} {\bibfield  {journal}
		{\bibinfo  {journal} {Nature Communications}\ }\textbf {\bibinfo {volume}
			{4}},\ \bibinfo {pages} {1913} (\bibinfo {year} {2013})}\BibitemShut
	{NoStop}%
	\bibitem [{\citenamefont {Gustavsson}\ \emph {et~al.}(2016)\citenamefont
		{Gustavsson}, \citenamefont {Yan}, \citenamefont {Catelani}, \citenamefont
		{Bylander}, \citenamefont {Kamal}, \citenamefont {Birenbaum}, \citenamefont
		{Hover}, \citenamefont {Rosenberg}, \citenamefont {Samach}, \citenamefont
		{Sears}, \citenamefont {Weber}, \citenamefont {Yoder}, \citenamefont
		{Clarke}, \citenamefont {Kerman}, \citenamefont {Yoshihara}, \citenamefont
		{Nakamura}, \citenamefont {Orlando},\ and\ \citenamefont
		{Oliver}}]{Gustavsson2016}%
	\BibitemOpen
	\bibfield  {author} {\bibinfo {author} {\bibfnamefont {S.}~\bibnamefont
			{Gustavsson}}, \bibinfo {author} {\bibfnamefont {F.}~\bibnamefont {Yan}},
		\bibinfo {author} {\bibfnamefont {G.}~\bibnamefont {Catelani}}, \bibinfo
		{author} {\bibfnamefont {J.}~\bibnamefont {Bylander}}, \bibinfo {author}
		{\bibfnamefont {A.}~\bibnamefont {Kamal}}, \bibinfo {author} {\bibfnamefont
			{J.}~\bibnamefont {Birenbaum}}, \bibinfo {author} {\bibfnamefont
			{D.}~\bibnamefont {Hover}}, \bibinfo {author} {\bibfnamefont
			{D.}~\bibnamefont {Rosenberg}}, \bibinfo {author} {\bibfnamefont
			{G.}~\bibnamefont {Samach}}, \bibinfo {author} {\bibfnamefont {A.~P.}\
			\bibnamefont {Sears}}, \bibinfo {author} {\bibfnamefont {S.~J.}\ \bibnamefont
			{Weber}}, \bibinfo {author} {\bibfnamefont {J.~L.}\ \bibnamefont {Yoder}},
		\bibinfo {author} {\bibfnamefont {J.}~\bibnamefont {Clarke}}, \bibinfo
		{author} {\bibfnamefont {A.~J.}\ \bibnamefont {Kerman}}, \bibinfo {author}
		{\bibfnamefont {F.}~\bibnamefont {Yoshihara}}, \bibinfo {author}
		{\bibfnamefont {Y.}~\bibnamefont {Nakamura}}, \bibinfo {author}
		{\bibfnamefont {T.~P.}\ \bibnamefont {Orlando}}, \ and\ \bibinfo {author}
		{\bibfnamefont {W.~D.}\ \bibnamefont {Oliver}},\ }\href {\doibase
		10.1126/science.aah5844} {\bibfield  {journal} {\bibinfo  {journal}
			{Science}\ }\textbf {\bibinfo {volume} {354}},\ \bibinfo {pages} {1573}
		(\bibinfo {year} {2016})}\BibitemShut {NoStop}%
	\bibitem [{\citenamefont {Schneider}\ \emph {et~al.}(2019)\citenamefont
		{Schneider}, \citenamefont {Wolz}, \citenamefont {Pfirrmann}, \citenamefont
		{Spiecker}, \citenamefont {Rotzinger}, \citenamefont {Ustinov},\ and\
		\citenamefont {Weides}}]{Schneider2019}%
	\BibitemOpen
	\bibfield  {author} {\bibinfo {author} {\bibfnamefont {A.}~\bibnamefont
			{Schneider}}, \bibinfo {author} {\bibfnamefont {T.}~\bibnamefont {Wolz}},
		\bibinfo {author} {\bibfnamefont {M.}~\bibnamefont {Pfirrmann}}, \bibinfo
		{author} {\bibfnamefont {M.}~\bibnamefont {Spiecker}}, \bibinfo {author}
		{\bibfnamefont {H.}~\bibnamefont {Rotzinger}}, \bibinfo {author}
		{\bibfnamefont {A.~V.}\ \bibnamefont {Ustinov}}, \ and\ \bibinfo {author}
		{\bibfnamefont {M.}~\bibnamefont {Weides}},\ }\href
	{https://arxiv.org/abs/1904.00208} {\bibfield  {journal} {\bibinfo  {journal}
			{arXiv}\ } (\bibinfo {year} {2019})}\BibitemShut {NoStop}%
	\bibitem [{\citenamefont {Kumar}\ \emph {et~al.}(2016)\citenamefont {Kumar},
		\citenamefont {Sendelbach}, \citenamefont {Beck}, \citenamefont {Freeland},
		\citenamefont {Wang}, \citenamefont {Wang}, \citenamefont {Yu}, \citenamefont
		{Wu}, \citenamefont {Pappas},\ and\ \citenamefont {McDermott}}]{Kumar2016}%
	\BibitemOpen
	\bibfield  {author} {\bibinfo {author} {\bibfnamefont {P.}~\bibnamefont
			{Kumar}}, \bibinfo {author} {\bibfnamefont {S.}~\bibnamefont {Sendelbach}},
		\bibinfo {author} {\bibfnamefont {M.}~\bibnamefont {Beck}}, \bibinfo {author}
		{\bibfnamefont {J.}~\bibnamefont {Freeland}}, \bibinfo {author}
		{\bibfnamefont {Z.}~\bibnamefont {Wang}}, \bibinfo {author} {\bibfnamefont
			{H.}~\bibnamefont {Wang}}, \bibinfo {author} {\bibfnamefont {C.~C.}\
			\bibnamefont {Yu}}, \bibinfo {author} {\bibfnamefont {R.}~\bibnamefont {Wu}},
		\bibinfo {author} {\bibfnamefont {D.}~\bibnamefont {Pappas}}, \ and\ \bibinfo
		{author} {\bibfnamefont {R.}~\bibnamefont {McDermott}},\ }\href {\doibase
		10.1103/physrevapplied.6.041001} {\bibfield  {journal} {\bibinfo  {journal}
			{Physical Review Applied}\ }\textbf {\bibinfo {volume} {6}} (\bibinfo {year}
		{2016}),\ 10.1103/physrevapplied.6.041001}\BibitemShut {NoStop}%
	\bibitem [{\citenamefont {Anton}\ \emph {et~al.}(2012)\citenamefont {Anton},
		\citenamefont {Nugroho}, \citenamefont {Birenbaum}, \citenamefont {O'Kelley},
		\citenamefont {Orlyanchik}, \citenamefont {Dove}, \citenamefont {Olson},
		\citenamefont {Yoscovits}, \citenamefont {Eckstein}, \citenamefont
		{Harlingen},\ and\ \citenamefont {Clarke}}]{Anton2012a}%
	\BibitemOpen
	\bibfield  {author} {\bibinfo {author} {\bibfnamefont {S.~M.}\ \bibnamefont
			{Anton}}, \bibinfo {author} {\bibfnamefont {C.~D.}\ \bibnamefont {Nugroho}},
		\bibinfo {author} {\bibfnamefont {J.~S.}\ \bibnamefont {Birenbaum}}, \bibinfo
		{author} {\bibfnamefont {S.~R.}\ \bibnamefont {O'Kelley}}, \bibinfo {author}
		{\bibfnamefont {V.}~\bibnamefont {Orlyanchik}}, \bibinfo {author}
		{\bibfnamefont {A.~F.}\ \bibnamefont {Dove}}, \bibinfo {author}
		{\bibfnamefont {G.~A.}\ \bibnamefont {Olson}}, \bibinfo {author}
		{\bibfnamefont {Z.~R.}\ \bibnamefont {Yoscovits}}, \bibinfo {author}
		{\bibfnamefont {J.~N.}\ \bibnamefont {Eckstein}}, \bibinfo {author}
		{\bibfnamefont {D.~J.~V.}\ \bibnamefont {Harlingen}}, \ and\ \bibinfo
		{author} {\bibfnamefont {J.}~\bibnamefont {Clarke}},\ }\href {\doibase
		10.1063/1.4749282} {\bibfield  {journal} {\bibinfo  {journal} {Applied
				Physics Letters}\ }\textbf {\bibinfo {volume} {101}},\ \bibinfo {pages}
		{092601} (\bibinfo {year} {2012})}\BibitemShut {NoStop}%
	\bibitem [{\citenamefont {Bruno}\ \emph {et~al.}(2015)\citenamefont {Bruno},
		\citenamefont {de~Lange}, \citenamefont {Asaad}, \citenamefont {van~der
			Enden}, \citenamefont {Langford},\ and\ \citenamefont {DiCarlo}}]{Bruno2015}%
	\BibitemOpen
	\bibfield  {author} {\bibinfo {author} {\bibfnamefont {A.}~\bibnamefont
			{Bruno}}, \bibinfo {author} {\bibfnamefont {G.}~\bibnamefont {de~Lange}},
		\bibinfo {author} {\bibfnamefont {S.}~\bibnamefont {Asaad}}, \bibinfo
		{author} {\bibfnamefont {K.~L.}\ \bibnamefont {van~der Enden}}, \bibinfo
		{author} {\bibfnamefont {N.~K.}\ \bibnamefont {Langford}}, \ and\ \bibinfo
		{author} {\bibfnamefont {L.}~\bibnamefont {DiCarlo}},\ }\href {\doibase
		10.1063/1.4919761} {\bibfield  {journal} {\bibinfo  {journal} {Applied
				Physics Letters}\ }\textbf {\bibinfo {volume} {106}},\ \bibinfo {pages}
		{182601} (\bibinfo {year} {2015})}\BibitemShut {NoStop}%
	\bibitem [{\citenamefont {Burnett}\ \emph {et~al.}(2019)\citenamefont
		{Burnett}, \citenamefont {Bengtsson}, \citenamefont {Scigliuzzo},
		\citenamefont {Niepce}, \citenamefont {Kudra}, \citenamefont {Delsing},\ and\
		\citenamefont {Bylander}}]{Burnett2019}%
	\BibitemOpen
	\bibfield  {author} {\bibinfo {author} {\bibfnamefont {J.~J.}\ \bibnamefont
			{Burnett}}, \bibinfo {author} {\bibfnamefont {A.}~\bibnamefont {Bengtsson}},
		\bibinfo {author} {\bibfnamefont {M.}~\bibnamefont {Scigliuzzo}}, \bibinfo
		{author} {\bibfnamefont {D.}~\bibnamefont {Niepce}}, \bibinfo {author}
		{\bibfnamefont {M.}~\bibnamefont {Kudra}}, \bibinfo {author} {\bibfnamefont
			{P.}~\bibnamefont {Delsing}}, \ and\ \bibinfo {author} {\bibfnamefont
			{J.}~\bibnamefont {Bylander}},\ }\href {\doibase 10.1038/s41534-019-0168-5}
	{\bibfield  {journal} {\bibinfo  {journal} {npj Quantum Information}\
		}\textbf {\bibinfo {volume} {5}} (\bibinfo {year} {2019}),\
		10.1038/s41534-019-0168-5}\BibitemShut {NoStop}%
	\bibitem [{\citenamefont {Schneider}\ \emph {et~al.}(2018)\citenamefont
		{Schneider}, \citenamefont {Braum{\"u}ller}, \citenamefont {Guo},
		\citenamefont {Stehle}, \citenamefont {Rotzinger}, \citenamefont {Marthaler},
		\citenamefont {Ustinov},\ and\ \citenamefont {Weides}}]{Schneider2018a}%
	\BibitemOpen
	\bibfield  {author} {\bibinfo {author} {\bibfnamefont {A.}~\bibnamefont
			{Schneider}}, \bibinfo {author} {\bibfnamefont {J.}~\bibnamefont
			{Braum{\"u}ller}}, \bibinfo {author} {\bibfnamefont {L.}~\bibnamefont {Guo}},
		\bibinfo {author} {\bibfnamefont {P.}~\bibnamefont {Stehle}}, \bibinfo
		{author} {\bibfnamefont {H.}~\bibnamefont {Rotzinger}}, \bibinfo {author}
		{\bibfnamefont {M.}~\bibnamefont {Marthaler}}, \bibinfo {author}
		{\bibfnamefont {A.~V.}\ \bibnamefont {Ustinov}}, \ and\ \bibinfo {author}
		{\bibfnamefont {M.}~\bibnamefont {Weides}},\ }\href {\doibase
		10.1103/physreva.97.062334} {\bibfield  {journal} {\bibinfo  {journal}
			{Physical Review A}\ }\textbf {\bibinfo {volume} {97}} (\bibinfo {year}
		{2018}),\ 10.1103/physreva.97.062334}\BibitemShut {NoStop}%
	\bibitem [{\citenamefont {Ithier}\ \emph {et~al.}(2005)\citenamefont {Ithier},
		\citenamefont {Collin}, \citenamefont {Joyez}, \citenamefont {Meeson},
		\citenamefont {Vion}, \citenamefont {Esteve}, \citenamefont {Chiarello},
		\citenamefont {Shnirman}, \citenamefont {Makhlin}, \citenamefont {Schriefl},\
		and\ \citenamefont {Schön}}]{Ithier2005}%
	\BibitemOpen
	\bibfield  {author} {\bibinfo {author} {\bibfnamefont {G.}~\bibnamefont
			{Ithier}}, \bibinfo {author} {\bibfnamefont {E.}~\bibnamefont {Collin}},
		\bibinfo {author} {\bibfnamefont {P.}~\bibnamefont {Joyez}}, \bibinfo
		{author} {\bibfnamefont {P.~J.}\ \bibnamefont {Meeson}}, \bibinfo {author}
		{\bibfnamefont {D.}~\bibnamefont {Vion}}, \bibinfo {author} {\bibfnamefont
			{D.}~\bibnamefont {Esteve}}, \bibinfo {author} {\bibfnamefont
			{F.}~\bibnamefont {Chiarello}}, \bibinfo {author} {\bibfnamefont
			{A.}~\bibnamefont {Shnirman}}, \bibinfo {author} {\bibfnamefont
			{Y.}~\bibnamefont {Makhlin}}, \bibinfo {author} {\bibfnamefont
			{J.}~\bibnamefont {Schriefl}}, \ and\ \bibinfo {author} {\bibfnamefont
			{G.}~\bibnamefont {Schön}},\ }\href {\doibase 10.1103/physrevb.72.134519}
	{\bibfield  {journal} {\bibinfo  {journal} {Physical Review B}\ }\textbf
		{\bibinfo {volume} {72}} (\bibinfo {year} {2005}),\
		10.1103/physrevb.72.134519}\BibitemShut {NoStop}%
	\bibitem [{\citenamefont {Luthi}\ \emph {et~al.}(2018)\citenamefont {Luthi},
		\citenamefont {Stavenga}, \citenamefont {Enzing}, \citenamefont {Bruno},
		\citenamefont {Dickel}, \citenamefont {Langford}, \citenamefont {Rol},
		\citenamefont {Jespersen}, \citenamefont {Nyg{\aa}rd}, \citenamefont
		{Krogstrup},\ and\ \citenamefont {DiCarlo}}]{Luthi2018}%
	\BibitemOpen
	\bibfield  {author} {\bibinfo {author} {\bibfnamefont {F.}~\bibnamefont
			{Luthi}}, \bibinfo {author} {\bibfnamefont {T.}~\bibnamefont {Stavenga}},
		\bibinfo {author} {\bibfnamefont {O.}~\bibnamefont {Enzing}}, \bibinfo
		{author} {\bibfnamefont {A.}~\bibnamefont {Bruno}}, \bibinfo {author}
		{\bibfnamefont {C.}~\bibnamefont {Dickel}}, \bibinfo {author} {\bibfnamefont
			{N.}~\bibnamefont {Langford}}, \bibinfo {author} {\bibfnamefont
			{M.}~\bibnamefont {Rol}}, \bibinfo {author} {\bibfnamefont {T.}~\bibnamefont
			{Jespersen}}, \bibinfo {author} {\bibfnamefont {J.}~\bibnamefont
			{Nyg{\aa}rd}}, \bibinfo {author} {\bibfnamefont {P.}~\bibnamefont
			{Krogstrup}}, \ and\ \bibinfo {author} {\bibfnamefont {L.}~\bibnamefont
			{DiCarlo}},\ }\href {\doibase 10.1103/physrevlett.120.100502} {\bibfield
		{journal} {\bibinfo  {journal} {Physical Review Letters}\ }\textbf {\bibinfo
			{volume} {120}} (\bibinfo {year} {2018}),\
		10.1103/physrevlett.120.100502}\BibitemShut {NoStop}%
	\bibitem [{\citenamefont {Welch}(1967)}]{Welch1967}%
	\BibitemOpen
	\bibfield  {author} {\bibinfo {author} {\bibfnamefont {P.}~\bibnamefont
			{Welch}},\ }\href {\doibase 10.1109/tau.1967.1161901} {\bibfield  {journal}
		{\bibinfo  {journal} {{IEEE} Transactions on Audio and Electroacoustics}\
		}\textbf {\bibinfo {volume} {15}},\ \bibinfo {pages} {70} (\bibinfo {year}
		{1967})}\BibitemShut {NoStop}%
	\bibitem [{\citenamefont {{F. J. Harris}}(1978)}]{Harris1978}%
	\BibitemOpen
	\bibfield  {author} {\bibinfo {author} {\bibnamefont {{F. J. Harris}}},\
	}\href {\doibase 10.1109/proc.1978.10837} {\bibfield  {journal} {\bibinfo
			{journal} {Proceedings of the {IEEE}}\ }\textbf {\bibinfo {volume} {66}},\
		\bibinfo {pages} {51} (\bibinfo {year} {1978})}\BibitemShut {NoStop}%
	\bibitem [{\citenamefont {Zeng}\ \emph {et~al.}(2015)\citenamefont {Zeng},
		\citenamefont {Nik}, \citenamefont {Greibe}, \citenamefont {Krantz},
		\citenamefont {Wilson}, \citenamefont {Delsing},\ and\ \citenamefont
		{Olsson}}]{Zeng2015}%
	\BibitemOpen
	\bibfield  {author} {\bibinfo {author} {\bibfnamefont {L.~J.}\ \bibnamefont
			{Zeng}}, \bibinfo {author} {\bibfnamefont {S.}~\bibnamefont {Nik}}, \bibinfo
		{author} {\bibfnamefont {T.}~\bibnamefont {Greibe}}, \bibinfo {author}
		{\bibfnamefont {P.}~\bibnamefont {Krantz}}, \bibinfo {author} {\bibfnamefont
			{C.~M.}\ \bibnamefont {Wilson}}, \bibinfo {author} {\bibfnamefont
			{P.}~\bibnamefont {Delsing}}, \ and\ \bibinfo {author} {\bibfnamefont
			{E.}~\bibnamefont {Olsson}},\ }\href {\doibase
		10.1088/0022-3727/48/39/395308} {\bibfield  {journal} {\bibinfo  {journal}
			{Journal of Physics D: Applied Physics}\ }\textbf {\bibinfo {volume} {48}},\
		\bibinfo {pages} {395308} (\bibinfo {year} {2015})}\BibitemShut {NoStop}%
	\bibitem [{Note1()}]{Note1}%
	\BibitemOpen
	\bibinfo {note} {A.Bilmes PhD thesis, Karlsruhe Institute of Technology,
		2019}\BibitemShut {NoStop}%
	\bibitem [{\citenamefont {Calusine}\ \emph {et~al.}(2018)\citenamefont
		{Calusine}, \citenamefont {Melville}, \citenamefont {Woods}, \citenamefont
		{Das}, \citenamefont {Stull}, \citenamefont {Bolkhovsky}, \citenamefont
		{Braje}, \citenamefont {Hover}, \citenamefont {Kim}, \citenamefont {Miloshi},
		\citenamefont {Rosenberg}, \citenamefont {Sevi}, \citenamefont {Yoder},
		\citenamefont {Dauler},\ and\ \citenamefont {Oliver}}]{Calusine2018}%
	\BibitemOpen
	\bibfield  {author} {\bibinfo {author} {\bibfnamefont {G.}~\bibnamefont
			{Calusine}}, \bibinfo {author} {\bibfnamefont {A.}~\bibnamefont {Melville}},
		\bibinfo {author} {\bibfnamefont {W.}~\bibnamefont {Woods}}, \bibinfo
		{author} {\bibfnamefont {R.}~\bibnamefont {Das}}, \bibinfo {author}
		{\bibfnamefont {C.}~\bibnamefont {Stull}}, \bibinfo {author} {\bibfnamefont
			{V.}~\bibnamefont {Bolkhovsky}}, \bibinfo {author} {\bibfnamefont
			{D.}~\bibnamefont {Braje}}, \bibinfo {author} {\bibfnamefont
			{D.}~\bibnamefont {Hover}}, \bibinfo {author} {\bibfnamefont {D.~K.}\
			\bibnamefont {Kim}}, \bibinfo {author} {\bibfnamefont {X.}~\bibnamefont
			{Miloshi}}, \bibinfo {author} {\bibfnamefont {D.}~\bibnamefont {Rosenberg}},
		\bibinfo {author} {\bibfnamefont {A.}~\bibnamefont {Sevi}}, \bibinfo {author}
		{\bibfnamefont {J.~L.}\ \bibnamefont {Yoder}}, \bibinfo {author}
		{\bibfnamefont {E.}~\bibnamefont {Dauler}}, \ and\ \bibinfo {author}
		{\bibfnamefont {W.~D.}\ \bibnamefont {Oliver}},\ }\href {\doibase
		10.1063/1.5006888} {\bibfield  {journal} {\bibinfo  {journal} {Applied
				Physics Letters}\ }\textbf {\bibinfo {volume} {112}},\ \bibinfo {pages}
		{062601} (\bibinfo {year} {2018})}\BibitemShut {NoStop}%
	\bibitem [{\citenamefont {Neill}\ \emph {et~al.}(2013)\citenamefont {Neill},
		\citenamefont {Megrant}, \citenamefont {Barends}, \citenamefont {Chen},
		\citenamefont {Chiaro}, \citenamefont {Kelly}, \citenamefont {Mutus},
		\citenamefont {O{\textquotesingle}Malley}, \citenamefont {Sank},
		\citenamefont {Wenner}, \citenamefont {White}, \citenamefont {Yin},
		\citenamefont {Cleland},\ and\ \citenamefont {Martinis}}]{Neill2013}%
	\BibitemOpen
	\bibfield  {author} {\bibinfo {author} {\bibfnamefont {C.}~\bibnamefont
			{Neill}}, \bibinfo {author} {\bibfnamefont {A.}~\bibnamefont {Megrant}},
		\bibinfo {author} {\bibfnamefont {R.}~\bibnamefont {Barends}}, \bibinfo
		{author} {\bibfnamefont {Y.}~\bibnamefont {Chen}}, \bibinfo {author}
		{\bibfnamefont {B.}~\bibnamefont {Chiaro}}, \bibinfo {author} {\bibfnamefont
			{J.}~\bibnamefont {Kelly}}, \bibinfo {author} {\bibfnamefont {J.~Y.}\
			\bibnamefont {Mutus}}, \bibinfo {author} {\bibfnamefont {P.~J.~J.}\
			\bibnamefont {O{\textquotesingle}Malley}}, \bibinfo {author} {\bibfnamefont
			{D.}~\bibnamefont {Sank}}, \bibinfo {author} {\bibfnamefont {J.}~\bibnamefont
			{Wenner}}, \bibinfo {author} {\bibfnamefont {T.~C.}\ \bibnamefont {White}},
		\bibinfo {author} {\bibfnamefont {Y.}~\bibnamefont {Yin}}, \bibinfo {author}
		{\bibfnamefont {A.~N.}\ \bibnamefont {Cleland}}, \ and\ \bibinfo {author}
		{\bibfnamefont {J.~M.}\ \bibnamefont {Martinis}},\ }\href {\doibase
		10.1063/1.4818710} {\bibfield  {journal} {\bibinfo  {journal} {Applied
				Physics Letters}\ }\textbf {\bibinfo {volume} {103}},\ \bibinfo {pages}
		{072601} (\bibinfo {year} {2013})}\BibitemShut {NoStop}%
	\bibitem [{\citenamefont {Gunnarsson}\ \emph {et~al.}(2013)\citenamefont
		{Gunnarsson}, \citenamefont {Pirkkalainen}, \citenamefont {Li}, \citenamefont
		{Paraoanu}, \citenamefont {Hakonen}, \citenamefont {Sillanpää},\ and\
		\citenamefont {Prunnila}}]{Gunnarsson2013}%
	\BibitemOpen
	\bibfield  {author} {\bibinfo {author} {\bibfnamefont {D.}~\bibnamefont
			{Gunnarsson}}, \bibinfo {author} {\bibfnamefont {J.-M.}\ \bibnamefont
			{Pirkkalainen}}, \bibinfo {author} {\bibfnamefont {J.}~\bibnamefont {Li}},
		\bibinfo {author} {\bibfnamefont {G.~S.}\ \bibnamefont {Paraoanu}}, \bibinfo
		{author} {\bibfnamefont {P.}~\bibnamefont {Hakonen}}, \bibinfo {author}
		{\bibfnamefont {M.}~\bibnamefont {Sillanpää}}, \ and\ \bibinfo {author}
		{\bibfnamefont {M.}~\bibnamefont {Prunnila}},\ }\href {\doibase
		10.1088/0953-2048/26/8/085010} {\bibfield  {journal} {\bibinfo  {journal}
			{Superconductor Science and Technology}\ }\textbf {\bibinfo {volume} {26}},\
		\bibinfo {pages} {085010} (\bibinfo {year} {2013})}\BibitemShut {NoStop}%
	\bibitem [{\citenamefont {Khalil}\ \emph {et~al.}(2014)\citenamefont {Khalil},
		\citenamefont {Gladchenko}, \citenamefont {Stoutimore}, \citenamefont
		{Wellstood}, \citenamefont {Burin},\ and\ \citenamefont
		{Osborn}}]{Khalil2014}%
	\BibitemOpen
	\bibfield  {author} {\bibinfo {author} {\bibfnamefont {M.~S.}\ \bibnamefont
			{Khalil}}, \bibinfo {author} {\bibfnamefont {S.}~\bibnamefont {Gladchenko}},
		\bibinfo {author} {\bibfnamefont {M.~J.~A.}\ \bibnamefont {Stoutimore}},
		\bibinfo {author} {\bibfnamefont {F.~C.}\ \bibnamefont {Wellstood}}, \bibinfo
		{author} {\bibfnamefont {A.~L.}\ \bibnamefont {Burin}}, \ and\ \bibinfo
		{author} {\bibfnamefont {K.~D.}\ \bibnamefont {Osborn}},\ }\href {\doibase
		10.1103/physrevb.90.100201} {\bibfield  {journal} {\bibinfo  {journal}
			{Physical Review B}\ }\textbf {\bibinfo {volume} {90}} (\bibinfo {year}
		{2014}),\ 10.1103/physrevb.90.100201}\BibitemShut {NoStop}%
	\bibitem [{\citenamefont {Nugroho}, \citenamefont {Orlyanchik},\ and\
		\citenamefont {Harlingen}(2013)}]{Nugroho2013}%
	\BibitemOpen
	\bibfield  {author} {\bibinfo {author} {\bibfnamefont {C.~D.}\ \bibnamefont
			{Nugroho}}, \bibinfo {author} {\bibfnamefont {V.}~\bibnamefont {Orlyanchik}},
		\ and\ \bibinfo {author} {\bibfnamefont {D.~J.~V.}\ \bibnamefont
			{Harlingen}},\ }\href {\doibase 10.1063/1.4801521} {\bibfield  {journal}
		{\bibinfo  {journal} {Applied Physics Letters}\ }\textbf {\bibinfo {volume}
			{102}},\ \bibinfo {pages} {142602} (\bibinfo {year} {2013})}\BibitemShut
	{NoStop}%
	\bibitem [{\citenamefont {Shalibo}\ \emph {et~al.}(2010)\citenamefont
		{Shalibo}, \citenamefont {Rofe}, \citenamefont {Shwa}, \citenamefont
		{Zeides}, \citenamefont {Neeley}, \citenamefont {Martinis},\ and\
		\citenamefont {Katz}}]{Shalibo2010}%
	\BibitemOpen
	\bibfield  {author} {\bibinfo {author} {\bibfnamefont {Y.}~\bibnamefont
			{Shalibo}}, \bibinfo {author} {\bibfnamefont {Y.}~\bibnamefont {Rofe}},
		\bibinfo {author} {\bibfnamefont {D.}~\bibnamefont {Shwa}}, \bibinfo {author}
		{\bibfnamefont {F.}~\bibnamefont {Zeides}}, \bibinfo {author} {\bibfnamefont
			{M.}~\bibnamefont {Neeley}}, \bibinfo {author} {\bibfnamefont {J.~M.}\
			\bibnamefont {Martinis}}, \ and\ \bibinfo {author} {\bibfnamefont
			{N.}~\bibnamefont {Katz}},\ }\href {\doibase 10.1103/physrevlett.105.177001}
	{\bibfield  {journal} {\bibinfo  {journal} {Physical Review Letters}\
		}\textbf {\bibinfo {volume} {105}} (\bibinfo {year} {2010}),\
		10.1103/physrevlett.105.177001}\BibitemShut {NoStop}%
	\bibitem [{\citenamefont {Wang}\ \emph {et~al.}(2009)\citenamefont {Wang},
		\citenamefont {Hofheinz}, \citenamefont {Wenner}, \citenamefont {Ansmann},
		\citenamefont {Bialczak}, \citenamefont {Lenander}, \citenamefont {Lucero},
		\citenamefont {Neeley}, \citenamefont {O'Connell}, \citenamefont {Sank},
		\citenamefont {Weides}, \citenamefont {Cleland},\ and\ \citenamefont
		{Martinis}}]{Wang2009}%
	\BibitemOpen
	\bibfield  {author} {\bibinfo {author} {\bibfnamefont {H.}~\bibnamefont
			{Wang}}, \bibinfo {author} {\bibfnamefont {M.}~\bibnamefont {Hofheinz}},
		\bibinfo {author} {\bibfnamefont {J.}~\bibnamefont {Wenner}}, \bibinfo
		{author} {\bibfnamefont {M.}~\bibnamefont {Ansmann}}, \bibinfo {author}
		{\bibfnamefont {R.~C.}\ \bibnamefont {Bialczak}}, \bibinfo {author}
		{\bibfnamefont {M.}~\bibnamefont {Lenander}}, \bibinfo {author}
		{\bibfnamefont {E.}~\bibnamefont {Lucero}}, \bibinfo {author} {\bibfnamefont
			{M.}~\bibnamefont {Neeley}}, \bibinfo {author} {\bibfnamefont {A.~D.}\
			\bibnamefont {O'Connell}}, \bibinfo {author} {\bibfnamefont {D.}~\bibnamefont
			{Sank}}, \bibinfo {author} {\bibfnamefont {M.}~\bibnamefont {Weides}},
		\bibinfo {author} {\bibfnamefont {A.~N.}\ \bibnamefont {Cleland}}, \ and\
		\bibinfo {author} {\bibfnamefont {J.~M.}\ \bibnamefont {Martinis}},\ }\href
	{\doibase 10.1063/1.3273372} {\bibfield  {journal} {\bibinfo  {journal}
			{Applied Physics Letters}\ }\textbf {\bibinfo {volume} {95}},\ \bibinfo
		{pages} {233508} (\bibinfo {year} {2009})}\BibitemShut {NoStop}%
	\bibitem [{\citenamefont {Chang}\ \emph {et~al.}(2013)\citenamefont {Chang},
		\citenamefont {Vissers}, \citenamefont {C{\'{o}}rcoles}, \citenamefont
		{Sandberg}, \citenamefont {Gao}, \citenamefont {Abraham}, \citenamefont
		{Chow}, \citenamefont {Gambetta}, \citenamefont {Rothwell}, \citenamefont
		{Keefe}, \citenamefont {Steffen},\ and\ \citenamefont {Pappas}}]{Chang2013}%
	\BibitemOpen
	\bibfield  {author} {\bibinfo {author} {\bibfnamefont {J.~B.}\ \bibnamefont
			{Chang}}, \bibinfo {author} {\bibfnamefont {M.~R.}\ \bibnamefont {Vissers}},
		\bibinfo {author} {\bibfnamefont {A.~D.}\ \bibnamefont {C{\'{o}}rcoles}},
		\bibinfo {author} {\bibfnamefont {M.}~\bibnamefont {Sandberg}}, \bibinfo
		{author} {\bibfnamefont {J.}~\bibnamefont {Gao}}, \bibinfo {author}
		{\bibfnamefont {D.~W.}\ \bibnamefont {Abraham}}, \bibinfo {author}
		{\bibfnamefont {J.~M.}\ \bibnamefont {Chow}}, \bibinfo {author}
		{\bibfnamefont {J.~M.}\ \bibnamefont {Gambetta}}, \bibinfo {author}
		{\bibfnamefont {M.~B.}\ \bibnamefont {Rothwell}}, \bibinfo {author}
		{\bibfnamefont {G.~A.}\ \bibnamefont {Keefe}}, \bibinfo {author}
		{\bibfnamefont {M.}~\bibnamefont {Steffen}}, \ and\ \bibinfo {author}
		{\bibfnamefont {D.~P.}\ \bibnamefont {Pappas}},\ }\href {\doibase
		10.1063/1.4813269} {\bibfield  {journal} {\bibinfo  {journal} {Applied
				Physics Letters}\ }\textbf {\bibinfo {volume} {103}},\ \bibinfo {pages}
		{012602} (\bibinfo {year} {2013})}\BibitemShut {NoStop}%
	\bibitem [{\citenamefont {Dynes}, \citenamefont {Narayanamurti},\ and\
		\citenamefont {Garno}(1978)}]{Dynes1978}%
	\BibitemOpen
	\bibfield  {author} {\bibinfo {author} {\bibfnamefont {R.~C.}\ \bibnamefont
			{Dynes}}, \bibinfo {author} {\bibfnamefont {V.}~\bibnamefont
			{Narayanamurti}}, \ and\ \bibinfo {author} {\bibfnamefont {J.~P.}\
			\bibnamefont {Garno}},\ }\href {\doibase 10.1103/physrevlett.41.1509}
	{\bibfield  {journal} {\bibinfo  {journal} {Physical Review Letters}\
		}\textbf {\bibinfo {volume} {41}},\ \bibinfo {pages} {1509} (\bibinfo {year}
		{1978})}\BibitemShut {NoStop}%
	\bibitem [{\citenamefont {Szab{\'{o}}}\ \emph {et~al.}(2016)\citenamefont
		{Szab{\'{o}}}, \citenamefont {Samuely}, \citenamefont {Ha{\v{s}}kov{\'{a}}},
		\citenamefont {Ka{\v{c}}mar{\v{c}}{\'{\i}}k}, \citenamefont
		{{\v{Z}}emli{\v{c}}ka}, \citenamefont {Grajcar}, \citenamefont {Rodrigo},\
		and\ \citenamefont {Samuely}}]{Szabo2016}%
	\BibitemOpen
	\bibfield  {author} {\bibinfo {author} {\bibfnamefont {P.}~\bibnamefont
			{Szab{\'{o}}}}, \bibinfo {author} {\bibfnamefont {T.}~\bibnamefont
			{Samuely}}, \bibinfo {author} {\bibfnamefont {V.}~\bibnamefont
			{Ha{\v{s}}kov{\'{a}}}}, \bibinfo {author} {\bibfnamefont {J.}~\bibnamefont
			{Ka{\v{c}}mar{\v{c}}{\'{\i}}k}}, \bibinfo {author} {\bibfnamefont
			{M.}~\bibnamefont {{\v{Z}}emli{\v{c}}ka}}, \bibinfo {author} {\bibfnamefont
			{M.}~\bibnamefont {Grajcar}}, \bibinfo {author} {\bibfnamefont {J.~G.}\
			\bibnamefont {Rodrigo}}, \ and\ \bibinfo {author} {\bibfnamefont
			{P.}~\bibnamefont {Samuely}},\ }\href {\doibase 10.1103/physrevb.93.014505}
	{\bibfield  {journal} {\bibinfo  {journal} {Physical Review B}\ }\textbf
		{\bibinfo {volume} {93}} (\bibinfo {year} {2016}),\
		10.1103/physrevb.93.014505}\BibitemShut {NoStop}%
	\bibitem [{\citenamefont {Herman}\ and\ \citenamefont
		{Hlubina}(2016)}]{Herman2016}%
	\BibitemOpen
	\bibfield  {author} {\bibinfo {author} {\bibfnamefont {F.}~\bibnamefont
			{Herman}}\ and\ \bibinfo {author} {\bibfnamefont {R.}~\bibnamefont
			{Hlubina}},\ }\href {\doibase 10.1103/physrevb.94.144508} {\bibfield
		{journal} {\bibinfo  {journal} {Physical Review B}\ }\textbf {\bibinfo
			{volume} {94}} (\bibinfo {year} {2016}),\
		10.1103/physrevb.94.144508}\BibitemShut {NoStop}%
\end{thebibliography}

\appendix

\section{\label{appendix:technical_details_appendix}Experimental details}

The sample was used in a previous publication~\cite{Schneider2018a}, and is described in detail there. The experimental environment is a liquid $\mathrm{^{4}He}$ cooled $^{3}\mathrm{He/^{4}He}$ dilution
refrigerator with a base temperature below $20\,\mathrm{mK}$ and thermal stability of $\pm1\,\mathrm{mK}$. The frequency of the readout resonator is $8.57\,\mathrm{GHz}$. Isolation of the qubit from the environment is ensured by $70\,\mathrm{dB}$ attenuation ($20\,\mathrm{dB}$ at $4\mathrm{\,K}$, $20\,\mathrm{dB}$ at $100\,\mathrm{mK}$, $20\,\mathrm{dB}$ at base temperature and $10\,\mathrm{dB}$ overall cable loss) from room temperature to the sample as well as two circulators at base temperature and a $6.3-15\,\mathrm{GHz}$ band pass filter placed before the first amplification stage at $4\mathrm{\,K}$. The sample was placed in a copper housing  in the first cooldown and an aluminum housing in the following runs, including the measurements presented in this paper. The sample was always enclosed by a Cryoperm magnetic shield. No systematic change in qubit parameters was observed for the two housing materials. Over all cooldowns, we observed $T_{1}$ times between $12$ and $\SI{80}{\micro\second}$, and $T_{2}^{\mathrm{R}}$ between $4$ and $\SI{90}{\micro\second}$.
\section{\label{appendix:measurement_details}Measurement details}

The reset time between measurements was chosen to be five times the longest observed mean relaxation time, resulting in a repetition rate of about $4\,\mathrm{kHz}$. Averaging was set according to the intended signal-to-noise ratio (SNR) and time resolution, usually between 200 to 1000 single shots were averaged for a single datapoint.

The maximum data acquisition rate is limited by the smallest number of points which still yield confident fits. The number of points required to characterize Ramsey oscillations depends on their frequency and decay time. A higher Ramsey frequency leads to  improved SNR for the frequency shift but requires a higher sampling rate. For a given number of points this implies a shorter interval of free evolution times, reducing the SNR of $T_{2}^{\mathrm{R}}$, leading to a tradeoff between good fits to the frequency shift
and Ramsey decay time for a given number of measurements. A high SNR is crucial for a conclusive PSD analysis, as statistical noise due to fit uncertainty raises the noise floor. We balance the distribution and number of points to achieve a tradeoff for the signal in different parameters. For example, the measurement depicted in Fig.~3 was optimized for accurate frequency fitting and achieves a mean error of $\pm 0.3\,\mathrm{kHz}$, but the mean dephasing-time error is $\pm\SI{10}{\micro\second}$. For comparison, the mean errors in Fig.~1\,(b) are $\pm\SI{2.7}{\micro\second}\,(T_1),\pm\SI{5.4}{\micro\second}\,(T_{2}^\mathrm{R})$, and $\pm\SI{0.9}{\kilo\hertz}\,(\Delta \omega_{\mathrm{q}})$.\\
Potential fluctuations of the readout resonator frequency $f_r$ only affect the SNR of our measurements but have no influence on the extracted parameters.

\section{\label{appendix:data_analysis_details}Data analysis details}

To minimize the effect of fit inaccuracy on our statistical analysis, e.g.\ due to fluctuations occurring during data acquisition for a single trace, or strong noise, unreliable fits with uncertainties larger than ten times the average are masked in the data sets. In the presented measurements, between 2 and $10\%$ of the single slices had inaccurate fitting and are not shown, the results are insensitive to the masking. \\We have verified that the detuning of pulses due to the measured shifts in qubit frequency does not lead to a systematic bias in the extracted parameters. For the observed fluctuation strength in frequency, the maximum change in signal amplitude of the decay curves is $0.8\,\%$.

Our extraction of the pure dephasing time $T_\mathrm{\Phi}$ associated with the rate $\Gamma_{\mathrm{\Phi}}$ implies a simple exponential decay in Ramsey measurements. While this is not necessarily the case \cite{Ithier2005}, the corresponding deviation compared to e.g.\ Gaussian decay is smaller than the fitting error. 

\begin{figure*}
	\includegraphics[width=0.9\textwidth]{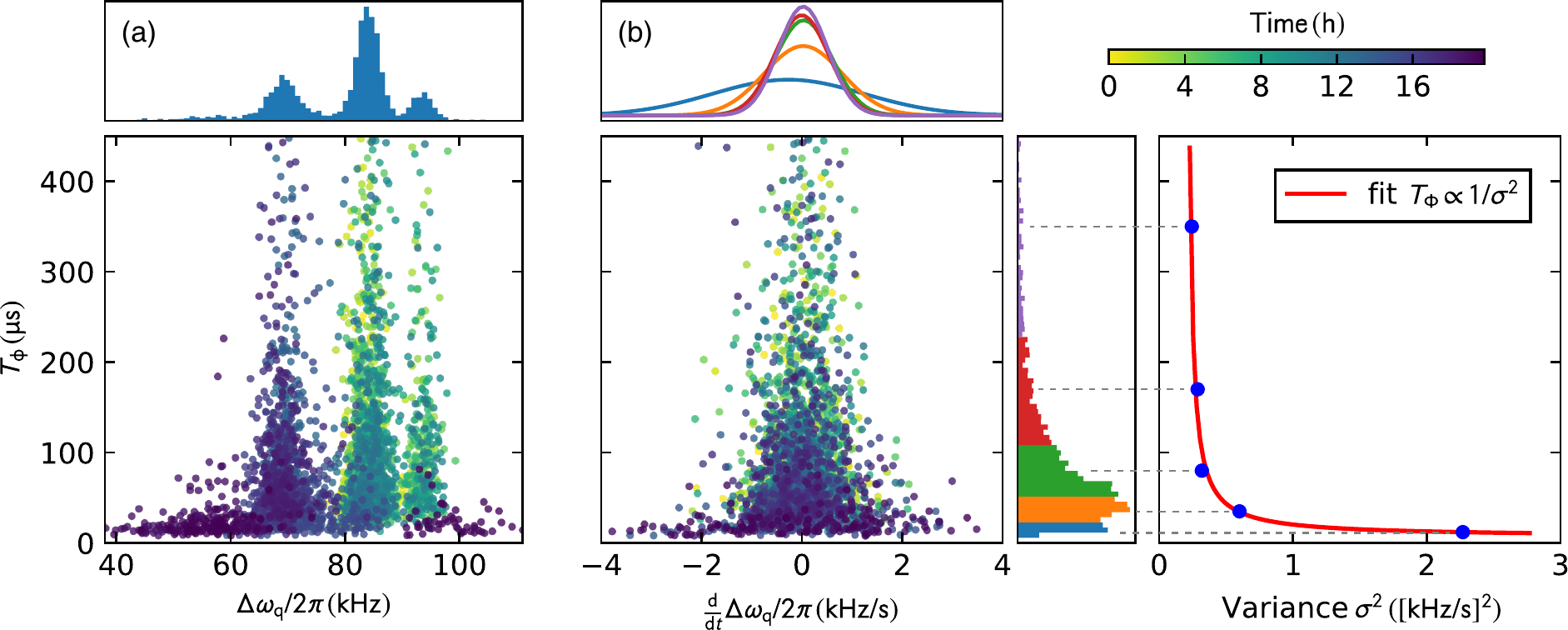}	
	\protect\caption{
		\label{fig:binned_scatter}Scatterplots of the pure dephasing time $T_{\Phi}$ versus Ramsey frequency shift (a) or fluctuation strength (b) of the data shown in Fig.\,1(b) in the main text. The point color indicates the measurement time. Discrete clusters in (a) are attributable to metastable switching of the qubit frequency. The smallest dephasing times correspond to the strongest spread in frequency. (b) The triangular shape in the scatterplot indicates that stronger fluctuations in $\Delta \omega_{\mathrm{q}}$ on the order of seconds are related to lower pure dephasing times, and was consistently measured in all cooldowns, regardless of fluctuation strength or coherence. Fits to normal distributions for different ranges of pure dephasing in the top panel show the increased variance in fluctuation strength at lower pure dephasing times. The functional dependence is roughly $T_\mathrm{\Phi}\propto 1/\sigma^2$. For visibility, the outliers due to switching which are symmetrically distributed around zero on the x-axis near $\pm 20\,\mathrm{kHz/s}$ are not shown.}
\end{figure*}

In addition to the discussion in the main text, we analyze the data of Fig.\,1(b) to characterize the dependence of fluctuation strength and pure dephasing time.
Scatterplots are shown in Fig.~\ref{fig:binned_scatter}, where the pure dephasing time $T_{\mathrm{\Phi}}$ is plotted either vs. Ramsey detuning $\Delta \omega_\mathrm{q}$ or the fluctuation strength $\frac{\mathrm{d}}{\mathrm{d}t}\,\Delta\omega_\mathrm{q}$. We attribute the bunching at discrete values of $\Delta \omega_\mathrm{q}$ visible in (a) to telegraphic spectral diffusion of near-resonant TLS, as also observed in other experiments~\cite{Lisenfeld2015,Klimov2018,Meisner2018,Luthi2018}.
In (b) we use the same method as in the main text. We bin the data  on the fluctuation strength according to their associated pure dephasing times (colored histogram), and fit the data in each bin to a Gaussian distribution (top panel). The corresponding variances $\sigma^2$ are plotted on the right and show increased variance for lower $T_{\mathrm{\Phi}}$, where the solid red line is a fit to  $T_\mathrm{\Phi}\propto 1/\sigma^2$. Normal distributions are chosen as we assume multiple independent sources contribute to the fluctuation and their errors. The fluctuation strength $\frac{\mathrm{d}}{\mathrm{d}t}\,\Delta\omega_\mathrm{q}$ is calculated as the difference of each data point to the previous divided by the elapsed time.

\section{\label{appendix:Power-Spectral-Density-Welch}Power Spectral Density estimation}

For PSD analysis we use Welch's method~\cite{Welch1967}. It achieves a reduction of noise by segmenting the data into smaller sets and sampling these with an overlapping window function in the time domain. The average squared magnitude of the discrete Fourier transforms of these samples then gives the power spectrum. This method reduces the frequency resolution but enhances the SNR of noise power measurements. Some details of the extracted PSD will, however depend on the window function and its size, which we provide in the following. In Fig.~3 we used a 'Kaiser' ($\alpha=4$) window~\cite{Harris1978}, with a segment size of 2844 points, corresponding to dividing the data into 10 samples, with $50\,\%$ overlap. Overall this set of data contained 14222 Ramsey measurements. All data processing has also been verified with random data, to exclude analysis artifacts from being identified as signal. The effective frequency limit and SNR of this measurement could be improved significantly, using a quantum-limited amplifier.

\section{Coupling strength and density estimation}
\label{appendix:coupling estimate}

Inside the qubit's Josephson junction, the coupling rate due to interaction of the TLS' dipole moment to the root mean square electric field of the qubits vacuum fluctuation is given by~\cite{Martinis2005} \begin{equation}\label{couplingstrength}
g_\mathrm{max}=\frac{|\vec{d}|}{2x}\sqrt{\frac{\omega_\mathrm{q}}{\pi h\,C_\mathrm{q}}}\approx 48\,\mathrm{MHz},
\end{equation} with the TLS' dipole moment $|\vec{d}|$ on the order of $1\,\mathrm{e\angstrom}$ ~\cite{Sarabi2016,Brehm2017,Lisenfeld2016}, the total qubit capacitance of $C_\mathrm{q}=120\,\mathrm{fF}$, $\omega_{\mathrm{q}}/2\pi=4.75\,\mathrm{GHz}$ and the width of the JJ capacitor $x$, corresponding to an estimated thickness of the oxide barrier of $1.8\,\mathrm{nm}$~\cite{Zeng2015}. In our experiment, the largest observed frequency fluctuations are $\SI{140}{\kilo \hertz}$. As the dispersive shift scales as $\chi = g^{2}/\Delta$, this implies a detuning between qubit and possible junction-TLS of the order of many $\mathrm{GHz}$. Alternating positive and negative correlation in Fig.~2\,(c) imply frequency diffusion across the qubit frequency, see also the raw data in Fig.~\ref{fig:Long-term-strong fluct}(a). Thermal switching of several $\mathrm{GHz}$ detuned TLS between below and above the qubit is unlikely. Therefore we assume the detuning between qubit and the most relevant (we call dominant) TLS to be close to zero for times of relatively large frequency shifts. If we assume the TLS to reside in the junction, the observed coupling implies their dipole moments to deviate less than \ang{0.1} from perpendicular to the qubits electric field, which is unlikely for several observed TLS. We reason it is unlikely that the observed frequency shifts are due to TLS which are located in the qubit's Josephson junction.

Using strain and electric field tuning, a recent work~\footnote{A.Bilmes PhD thesis, Karlsruhe Institute of Technology, 2019} possibly found junction TLS with much smaller dipole moments on the order of $1\,\mathrm{me\angstrom}$. TLS of such unusually small dipole moments in the junction could also account for the observed fluctuations. However, without further information we interpret our findings using standard values for dipole moments of TLS. 

Significant changes in quality factors of superconducting resonators due to surface treatment observed e.g.\ in Ref~\cite{Calusine2018} imply considerable coupling to surface TLS. The electric field at edges of metal films scales approximately as $1/\sqrt{x}$~\cite{Neill2013}. In our case, the resulting field strengths are larger than $4\,\mathrm{V/m}$  at any position closer than $20\,\mathrm{nm}$ to the superconducting film edges. For the typical TLS dipole moments of $d\approx 1\,\mathrm{e\angstrom}$, this corresponds to a coupling rate to the qubit of $g\gtrsim100\,\mathrm{kHz}$ in agreement with our observed qubit frequency shifts. The TLS mostly responsible for decoherence and frequency shifts are therefore presumably at the surface, near edges which create field enhancement. From simulations of the field distribution of our circuit, we know that the field drops to about $55\,\mathrm{mV/m}$ in the middle between the conductors, resulting in a maximum coupling rate of about $g=1.3\,\mathrm{kHz}$ there. This result again matches well with the smallest observed fluctuations of about $1\,\mathrm{kHz}$ and we conclude that our experiment is sensitive to TLS positioned anywhere in the shunt capacitance.

Although our fixed frequency qubit limits the accessible information on TLS density, we can deduce a rough estimate for the surface density of the dominant TLS based on the statistics of several measurements. We observed TLS with coupling rates on the order of $g_\mathrm{max} \approx 100\,\mathrm{kHz}$, and the observed frequency fluctuations for times of relatively high coherence and frequency stability are on the order of $\Delta \omega_\mathrm{q,min}/2\pi \approx 1\,\mathrm{kHz}$. Thus, assuming TLS of the coupling strength $g_\mathrm{max}$ are also present in times of high coherence, the detuning to such a TLS is approximately $2\pi\,g_\mathrm{max}^2/\Delta \omega_\mathrm{q,min} = 10\,\mathrm{MHz}$. At this frequency spacing of $20\,\mathrm{MHz}$, the frequency density of dominant TLS is $50/\mathrm{GHz}$. The surface area of our chip which during operation hosts electric field strengths larger than $4\,\mathrm{V/m}$ is about $\SI{164}{\micro\meter^2}$. This results in an estimated dominant-TLS surface density of $0.3/\mathrm{GHz}\,\SI{}{\micro\meter^2}$.\\
For comparison, the TLS densities reported by other groups are e.g.\ $0.5/\mathrm{GHz}\,\SI{}{\micro\meter^2}$ for large ($\approx1\,\SI{}{\micro\meter^2})$  $\mathrm{Al/AlO_x}$ junctions \cite{Martinis2005} or $2.4/\mathrm{GHz}\,\SI{}{\micro\meter^2}$ for significantly less coherent qubits \cite{Gunnarsson2013}. The low loss material $\mathrm{Si_3N_4}$  showed only $0.03/\mathrm{GHz}\,\SI{}{\micro\meter^2}$ in measurements on lumped-element resonators \cite{Khalil2014}. Compared to these values, our estimated TLS density is plausible and as expected smaller than for larger $\mathrm{Al/AlO_x}$ junction qubits.

\section{Cross-correlation specifics}
\label{appendix:correlation specifics}

Further cross-correlation analysis confirms the relationship between the observed frequency fluctuations and dephasing in all datasets. We conclude that the same mechanism is responsible for qubit dephasing and slow fluctuations of its parameters. Further, we gain insight into TLS-induced qubit relaxation. For periods of low pure dephasing, the absolute fluctuation strength and the relaxation rate $[|\frac{\mathrm{d}}{\mathrm{d}t}\,\Delta \omega_{\mathrm{q}}| \star\,\Gamma_{1}](t=0)$ correlate significantly (see Fig.~\ref{fig:Long-term-strong fluct}(b)). 
This shows that TLS can be a dominant photon loss channel in our system and renders alternative models as primary decoherence mechanism unlikely. For example critical current fluctuations generate similar noise spectra~\cite{Nugroho2013}, these however only affect the qubit frequency $\omega_{\mathrm{q}}$ 
and thus do not explain a correlation of $|\frac{\mathrm{d}}{\mathrm{d}t}\,\Delta \omega_{\mathrm{q}}|$ with $\Gamma_{1}$. 
The lifetime of individual high frequency TLS in $\mathrm{AlO_{x}}$ was found to range from nanoseconds to microseconds~\cite{Lisenfeld2016,Shalibo2010}. Thus, for qubits with several microsecond relaxation times like our sample, these TLS represent a relevant photon loss channel.

Assuming that a single dominant TLS is responsible for dephasing and photon loss implies that the cross correlations $\Delta \omega_{\mathrm{q}} \star \Gamma_{\mathrm{\Phi}}$ and $\Delta \omega_{\mathrm{q}} \star \Gamma_{1}$ have the same sign at zero delay.
While this was our typical observation, also the opposite behavior was encountered (see Fig.~\ref{fig:Long-term-strong fluct}(b)), pointing towards different sources for dephasing and relaxation in those cases. 
This can be explained by the presence of a stable TLS 'A' close to resonance with the qubit, increasing its relaxation, and a second more weakly coupled TLS 'B' which fluctuates and is mainly responsible for dephasing. Correlations of opposite sign emerge if both TLS are above or below the qubit frequency. 
In that case, diffusion of B towards the qubit frequency results in level repulsion, detuning the qubit further from A. 

During periods of high coherence, we observe no cross-correlation between $\Gamma_{1}$ and $|\frac{\mathrm{d}}{\mathrm{d}t}\,\Delta \omega_{\mathrm{q}}|$ (see Fig.~\ref{fig:Long-term-no fluct}(b)). We interpret this, as due to a bath of weakly coupled TLS limiting $\Gamma_{1}$, rather than a single strongly coupled TLS~\cite{Wang2009,Chang2013,Calusine2018}. In the same measurement we still observed some correlation of the absolute fluctuation strength with $\Gamma_{\Phi}$. Our explanation is the different scaling of relaxation $\Gamma_{1}\propto g^2/\Delta^2$ and dispersive shift $\chi_{k}\propto g^2/\Delta$. Thus, a TLS can be detuned far enough to still cause dephasing, but not dominate the relaxation. This would also explain the stronger fluctuations in $\Gamma_{\mathrm{\Phi}}$ compared to $\Gamma_{1}$, observed throughout our measurements.

\section{Other decoherence sources}
\label{appendix:dynes}
Increased quasiparticle density due to pair-breaking by in-gap states, as observed in $\mathrm{Pb_{0.9}Bi_{0.1}}$ junctions~\cite{Dynes1978} or thin MoC films~\cite{Szabo2016,Herman2016}, could increase the qubit's relaxation rate. To our knowledge, no evidence of excessive pair breaking due to in-gap states has been reported for material systems like our sample at $\SI{20}{\milli\kelvin}$. 
The expected effect of increased quasiparticle density due to inelastic scattering in the junction, would be a reduction in T1. But we see no possibility to explain the observed discrete fluctuations between metastable states in qubit frequency with in-gap states.

\section{Supplementary data}
\label{appendix:supplementary data}

\begin{figure*}
	\includegraphics[width=1\textwidth]{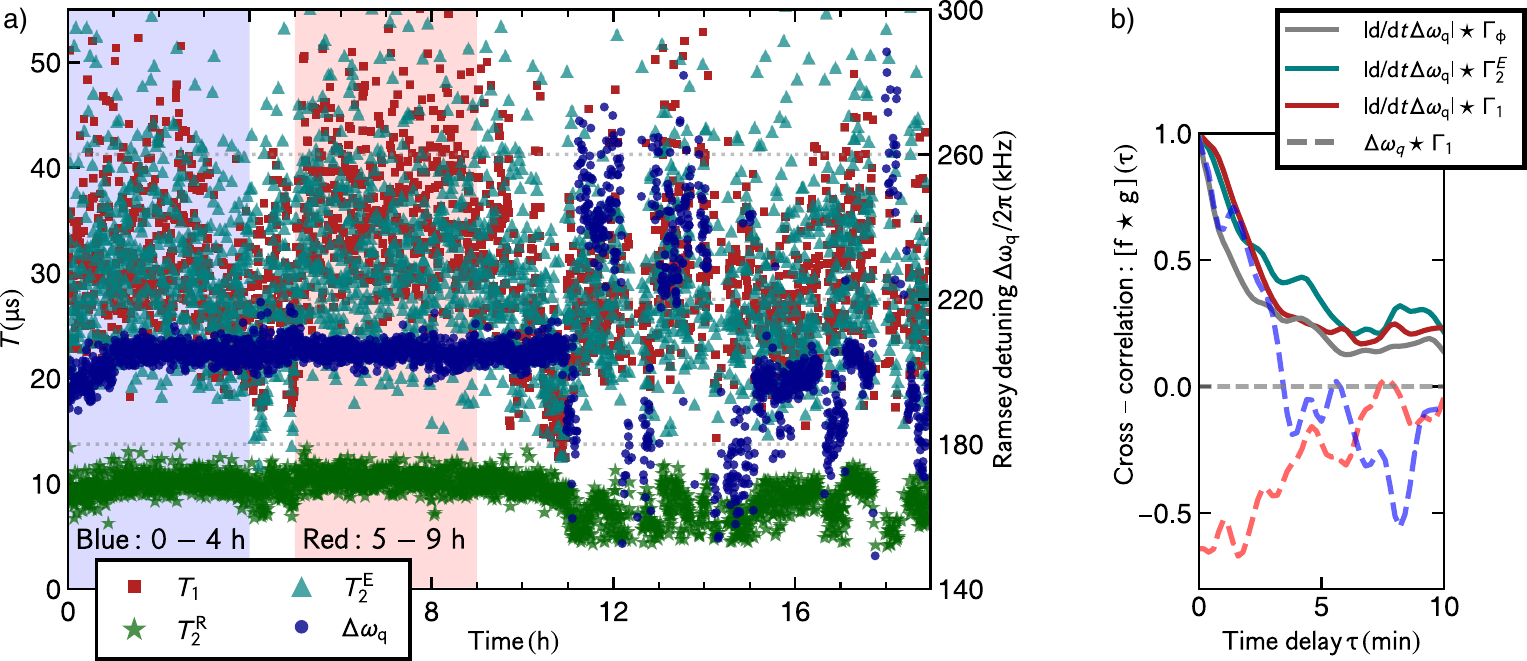}
	\protect\caption{
		\label{fig:Long-term-strong fluct} Data referred to in Fig.~2\,(a,b) with pronounced frequency changes of about $100\,\mathrm{kHz}$ and Ramsey dephasing times consistently below $\SI{15}{\micro\second}$ even during times without relatively strong fluctuation (first $10\,\mathrm{h}$). (b) Cross-correlations of the fluctuation strength or the qubit frequency with the rates of pure dephasing (gray), spin echo dephasing (teal), and relaxation (red). In addition to Fig.\,2(a) in the main text, the blue and red dashed lines represent the cross-correlation for the shaded areas of four hours respectively and for clarity, a weak smoothing is applied. At zero time delay $\tau$, significant correlation between the absolute fluctuation strength and $\Gamma_{\mathrm{\Phi}}$, $\Gamma_{2}^{E}$ and $\Gamma_{1}$ can be seen. The cross-correlation of $\Delta \omega_{\mathrm{q}}$ with $\Gamma_{1}$ (dashed lines) can change its sign for different intervals. This change in correlation was already seen for the pure dephasing in Fig.\,2(b). We interpret this as the result of spectral diffusion of a TLS crossing the qubit frequency.}    
\end{figure*}

\begin{figure*}
	\includegraphics[width=0.95\textwidth]{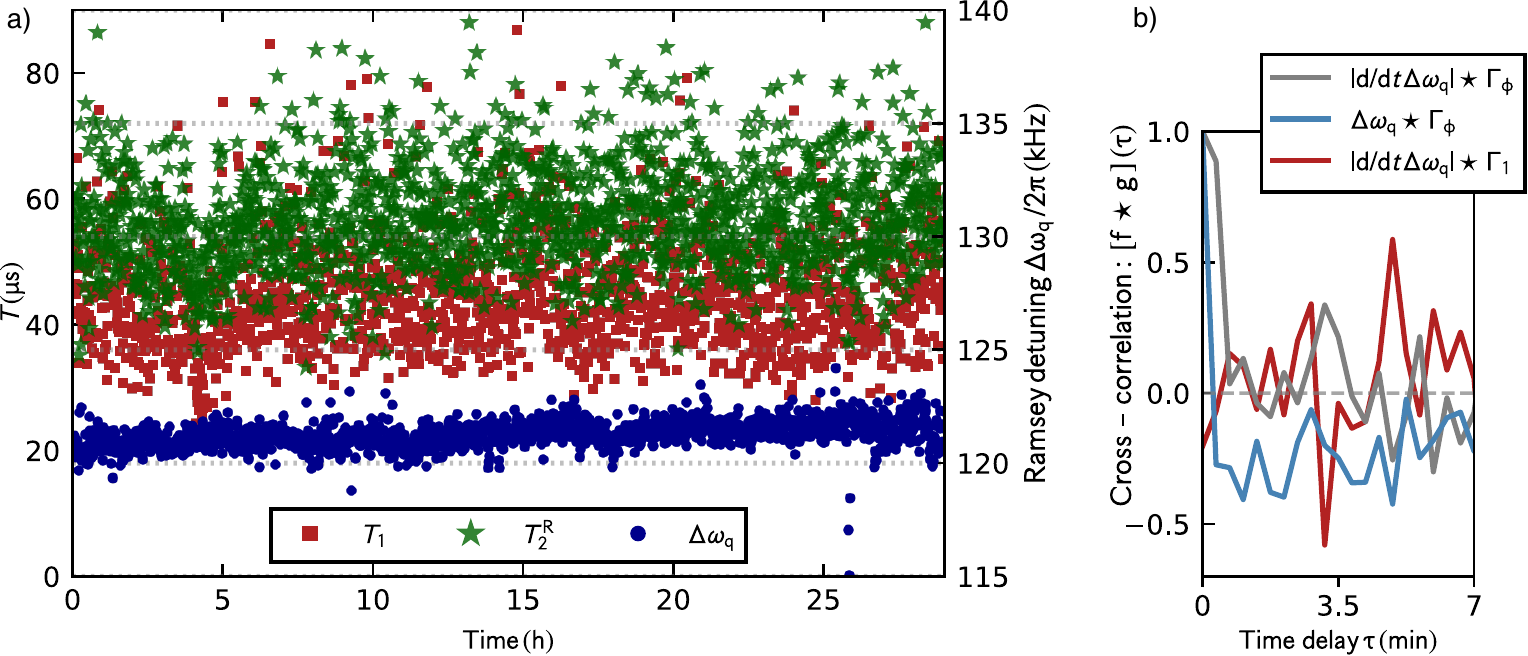}	
	\protect\caption{\label{fig:Long-term-no fluct}Subsequent cooldown with
		respect to Fig.~\ref{fig:Long-term-strong fluct} (no changes to the setup). The qubit frequency is relatively stable (mean frequency noise of $2\,\mathrm{kHz}$) and shows consistently high relaxation and dephasing times. A slow drift in frequency can be seen. (b) At zero time delay $\tau$, only $\Gamma_{\mathrm{\Phi}}$ shows a small correlation (compared to the noise-level) with the absolute fluctuation strength (gray) and the qubit frequency (light blue). No correlation with the relaxation rate was observed. }
\end{figure*}

Additional data referred to in the main text is shown in figures \ref{fig:Long-term-strong fluct} and \ref{fig:Long-term-no fluct}. About one year after fabrication, the sample was repeatedly measured over two years, no aging effects could be identified. We find the same pattern as in Fig.\,2(c) in the main text for all clusters representing metastable states as can be seen in Fig.~\ref{fig:binned_scatter}(a). 

\end{document}